\documentclass[journal]{IEEEtran}
\usepackage[top=0.7in, bottom=0.98in, left=0.7in, right=0.7in]{geometry}
\IEEEoverridecommandlockouts
\usepackage{cite}
\usepackage{subfig}
\usepackage{amsmath,amssymb,amsfonts}
\usepackage{algorithmic}
\usepackage{graphicx}
\usepackage{textcomp}
\usepackage{xcolor}
\usepackage{graphicx}
\usepackage{float}
\usepackage{bm}
\newtheorem{remark}{\rm{\textbf{Remark}}}
\newtheorem{proposition}{\rm{\textbf{Proposition}}}

\newtheorem{theorem}{\rm{\textbf{Theorem}}}
\usepackage{booktabs}   
\usepackage{multirow}   
\usepackage{caption}
\DeclareSubrefFormat{parens}{#1(#2)}
\def\BibTeX{{\rm B\kern-.05em{\sc i\kern-.025em b}\kern-.08em
    T\kern-.1667em\lower.7ex\hbox{E}\kern-.125emX}}
\usepackage{cmap} 
\usepackage[T1]{fontenc}
\usepackage{times} 
\usepackage[utf8]{inputenc}
\usepackage{threeparttable} 
\makeatother

\begin{document}

\title{Generalizable Learning for Massive MIMO CSI Feedback in Unseen Environments}
\vspace{-20pt}
\author{Haoyu Wang,~\IEEEmembership{Graduate Student Member,~IEEE,}~Zhi Sun,~\IEEEmembership{Senior Member,~IEEE,} \\
Shuangfeng Han,~\IEEEmembership{Senior Member,~IEEE,}~Xiaoyun Wang,~and Zhaocheng Wang,~\IEEEmembership{Fellow,~IEEE}

\vspace{-20pt}
    \thanks{Haoyu Wang, Zhi Sun, and Zhaocheng Wang are with the Department of Electronic Engineering, Tsinghua University, Beijing 100084 China (e-mail: wanghy22@mails.tsinghua.edu.cn; zhisun@ieee.org; zcwang@tsinghua.edu.cn).}
    \thanks{Shuangfeng Han and Xiaoyun Wang are with the China Mobile Research Institute, Beijing 100053, China. (e-mail: hanshuangfeng@chinamobile.com; wangxiaoyun@chinamobile.com)}
    \thanks{This work was supported by the National Key R\&D Program of China under Grant 2022YFB2902004. A shorter version has been accepted by IEEE GLOBECOM 2025 \cite{wang2025feedback}.}
    \thanks{Corresponding Author: Zhi Sun.}
}

\maketitle
\vspace{-20pt}
\begin{abstract}
Deep learning is promising to enhance the accuracy and reduce the overhead of channel state information (CSI) feedback, which can boost the capacity of frequency division duplex (FDD) massive multiple-input multiple-output (MIMO) systems. Nevertheless, the generalizability of current deep learning-based CSI feedback algorithms cannot be guaranteed in unseen environments, which induces a high deployment cost. In this paper, the generalizability of deep learning-based CSI feedback is promoted with physics interpretation. Firstly, the distribution shift of the cluster-based channel is modeled, which comprises the multi-cluster structure and single-cluster response. Secondly, the physics-based distribution alignment is proposed to effectively address the distribution shift of the cluster-based channel, which comprises multi-cluster decoupling and fine-grained alignment. Thirdly, the efficiency and robustness of physics-based distribution alignment are enhanced. Explicitly, an efficient multi-cluster decoupling algorithm is proposed based on the Eckart–Young-Mirsky (EYM) theorem to support real-time CSI feedback. Meanwhile, a hybrid criterion to estimate the number of decoupled clusters is designed, which enhances the robustness against channel estimation error. Fourthly, environment-generalizable neural network for CSI feedback (EG-CsiNet) is proposed as a novel learning framework with physics-based distribution alignment. Based on extensive simulations and sim-to-real experiments in various conditions, the proposed EG-CsiNet can robustly reduce the generalization error by more than 3 dB compared to the state-of-the-arts. 
\end{abstract}


\section{Introduction}
In the 5G and beyond 5G (B5G) wireless networks, massive multiple-input multiple-output (MIMO) is a pivotal technology to enhance the spectral efficiency (SE) and enable massive connectivity \cite{jsac_jin_2023_massive}. Benefiting from the large-scale antenna array, the diversity and multiplexing gain of massive MIMO systems can be significantly enhanced with the beamforming and precoding operations. To maximize the performance of beamforming and precoding, accurate downlink channel state information (CSI) at the BS is essential. In frequency division duplex (FDD) massive MIMO systems, the reciprocity between the uplink and downlink channels is not held due to the non-overlapping frequency bands \cite{gc_zhong_2020_fdd}. Thus, the acquisition of downlink CSI comprises two consecutive procedures, where the downlink channel is firstly estimated at the user and then fed back to the BS. Intuitively, the dimensions of the CSI matrix are proportional to the number of antennas and subcarriers, which induces a large feedback overhead for massive MIMO systems. Thus, accurate and low-overhead CSI feedback is vital to enhance the effective throughput of the massive MIMO systems. 

Fortunately, the wireless channel exhibits a sparse nature due to the limited scatterers in the propagation environment \cite{tsp_rao_2014_distributed,twc_wang_2023_squint}, which facilitates efficient CSI feedback with low overhead. Conventionally, compressed sensing (CS) \cite{tsp_rao_2014_distributed} and codebook-based \cite{3gpp.38.214} schemes are adopted to reduce the CSI feedback overhead. However, the CS-based CSI feedback relies on the strong assumption of a pre-defined sparse structure of the CSI matrix. Consequently, the accuracy of the reconstructed channel will be severely degraded when the pre-defined sparse structure is not met \cite{wcl_wen_2018_csinet}. In the codebook-based CSI feedback, the dominant ports are selected from the codebook and fed back to the BS. Nevertheless, correlations between ports are not utilized in the codebook-based CSI feedback, which limits the compression capability \cite{icc_sang_2024_type2}.

Compared to the conventional schemes, deep learning exhibits powerful compression capability in modeling the correlations in CSI, which facilitates low-overhead and accurate CSI feedback \cite{wcl_wen_2018_csinet,twc_guo_2020_csinetp,wcl_cui_2022_transnet}. 
Typically, the autoencoder (AE) is widely adopted in current deep learning-based CSI feedback, which can generate a low-dimensional codeword from the CSI. To better capture the sparse nature of CSI, numerous types of neural network (NN) structures have been proposed in the encoder and decoder modules, including the convolutional neural network (CNN)-based CsiNet/CsiNet+ \cite{wcl_wen_2018_csinet,twc_guo_2020_csinetp} and the transformer-based TransNet \cite{wcl_cui_2022_transnet}.

\begin{figure*}[t]
    \vspace{-5pt}
        \centering
        \includegraphics[width=1\textwidth]{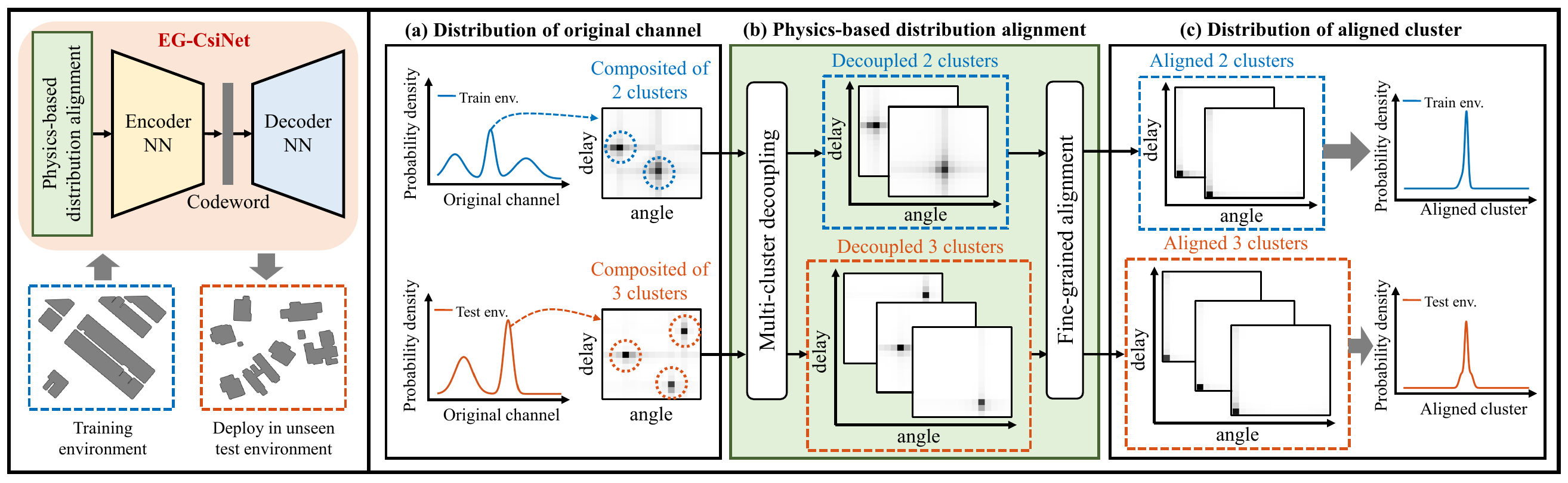}
    \captionsetup{font=footnotesize}
    \caption{Proposed EG-CsiNet with strong environment generalizability (left). The physics-based distribution alignment module in EG-CsiNet can effectively address the distribution shift of the cluster-based channel between training and test environments (right).}
    \label{fig: framework}
    \vspace{-15pt}
\end{figure*}

Despite the powerful compression capability, current deep learning-based CSI feedback algorithms exhibit limited generalizability to unseen environments, which has gained the attention of both academia \cite{tcom_guo_2022_overview} and industry \cite{3gpp.38.843}. Conventionally, CSI samples used for training and testing the autoencoders are drawn from the same distribution. However, in the practical deployment scenarios of massive MIMO systems, the distribution of CSI is environment-dependent, which is determined by the propagation conditions of the wireless environment. Consequently, the drastic CSI distribution shift frequently occurs due to the diverse electromagnetic waves propagation conditions. Thus, the out-of-distribution (OOD) generalizability of the pretrained model cannot be guaranteed in unseen environments \cite{comst_Akrout_2023_domain}, which poses challenges to the large-scale deployment. 

Model adaptation has been widely explored to improve the performance of deep learning-based CSI feedback under distribution shift between environments. These approaches typically assume that CSI samples from the target environment are available for fine-tuning or retraining the neural networks. Typical model adaptation schemes include transfer/meta-learning \cite{tgcn_zeng_2021_transfer,twc_han_2024_meta} and scenario-adaptive plugin design \cite{twc_liu_2024_deep}. In transfer/meta-learning \cite{tgcn_zeng_2021_transfer}, the parameters of the pretrained model are directly updated with the samples from the new environment. Additionally, a knowledge-driven meta-learning scheme has been proposed, where the samples from the target environments are augmented based on the statistical knowledge of the channel feature \cite{twc_han_2024_meta}. Further, a scenario-adaptive plug-in translation module is proposed \cite{twc_liu_2024_deep}, which is light-weighted and enables the reuse of the pretrained model. While these methods have shown effectiveness in improving environment-specific performance, they inherently rely on target-domain CSI samples for adaptation, which may not always be accessible in real-time or unseen environments. Consequently, their applicability in practical massive MIMO systems is limited, especially for online inference where no target-domain fine-tuning or retraining is feasible \cite{comst_Akrout_2023_domain}.

Environment-generalizable learning aims to directly deploy the pretrained model in new environments. Compared to the aforementioned model adaptation, the environment generalizable learning is free of CSI samples from new environments for fine-tuning or retraining, which can greatly reduce the deployment cost and enhance the real-time applicability \cite{comst_Akrout_2023_domain}. Since the CSI samples from the target environment are not accessible, environment-generalizable learning is conceptually more challenging than the model adaptation techniques, and the related works are limited. Current state-of-the-art (SOTA) environment generalizable learning algorithms for CSI feedback include the dataset-mixing \cite{twc_jiang_2024_multidomain} and UniversalNet/UniversalNet+ \cite{wcnc_liu_2024_generalization}. In the dataset-mixing \cite{twc_jiang_2024_multidomain}, a mixed dataset collected from multiple environments or sources is yielded for model pretraining.  However, the generalizability of the model is limited by the diversity of the mixed dataset since no specialized generalization modules are designed. In UniversalNet/UniversalNet+ \cite{wcnc_liu_2024_generalization}, the CSI samples are aligned with a benchmark via cross-correlation, which can reduce the inter-environment gap. However, the cross-correlation-based preprocessing is generally coarse, which does not leverage the inherent physics structure in the channel sample. Meanwhile, it is challenging to achieve fine-grained alignment to a single benchmark in the complex and diverse multipath environments.

In this paper, a novel environment-generalizable neural network for CSI feedback (EG-CsiNet) with intuitive physics interpretability is proposed, which is illustrated on the left of Fig.~\ref{fig: framework}. Firstly, the cross-environment distribution shift of the cluster-based channel is modeled, which can better reflect the realastic channel behaviour and reveal the inherent physics structure to understand the environment-generalizability of deep learning-based CSI feedback. When the objects in the wireless environments have rough surfaces or are densely distributed near the users, the propagated paths in the channel exhibit a clustered structure with similar delay and angle parameters. Thus, the original channel is composed of clusters and can be found in block (a) of Fig.~\ref{fig: framework}. Here, the probability distribution functions (PDFs) plot the distribution of the original channel, not the power profile of an individual channel. By comparing the PDFs, the misalignemnt of PDF curves indicates the channel distribution shift, including cluster number, multi-cluster dependency, and single-cluster response. Secondly, a physics-based distribution alignment is designed to effectively address the distribution shift of cluster-based channels, which comprises two modules of multi-cluster decoupling and fine-grained alignment. As shown in block (b) of Fig.~\ref{fig: framework}, the original channel is decoupled as the summation of individual clusters by applying multi-cluster decoupling, which addresses the distribution shift of cluster number and multi-cluster dependency. Then, fine-grained alignment is applied to each decoupled cluster to yield aligned clusters. As shown in block (c) of Fig.~\ref{fig: framework}, the distribution of the aligned cluster is stable across environments, where the distribution shift of single-cluster response is further addressed. Thirdly, the practical implementations of the physics-based distribution alignment are proposed. Specifically, cluster-level properties of the massive MIMO channel are derived. Then, an efficient singular value decomposition (SVD)-based multi-cluster decoupling is proposed based on the derived cluster-level properties and the Eckart-Young-Mirsky (EYM) theorem \cite{eckart1936approximation,mirsky1960symmetric}. Compared to the conventional algorithm \cite{wang2025generalizable}, the proposed SVD-based multi-cluster decoupling avoids the intermediate path-level parameter estimation, which reduces the computation complexity and facilitates the real-time application. Meanwhile, a hybrid criterion is proposed to estimate the number of clusters, which not only enhances noise robustness but also facilitates the compression capability for the unsupervised CSI feedback task. Fourthly, the EG-CsiNet for CSI feedback is proposed, where the model training and inference are designed. Explicitly, the neural networks in EG-CsiNet are trained to compress the distributional stable aligned clusters while the distributional varying components are processed by learning-free methods, which directly leverages the physics-based distribution alignment to guarantee model generalizability. Meanwhile, the feedback overhead and the model parameter complexity of EG-CsiNet are rigorously analyzed. On the one hand, the feedback overhead of the proposed EG-CsiNet can be adaptively adjusted based on the number of decoupled clusters. On the other hand, the model parameter in EG-CsiNet can also be reduced under the same feedback overhead. Based on extensive simulations, the proposed EG-CsiNet can robustly reduce the OOD generalization error by more than 3 dB compared to the SOTA. Meanwhile, the effectiveness of the proposed EG-CsiNet is robustly held in a challenging sim-to-real experiment, which further validates its potential for practical deployments.

The present work shares with our previous works \cite{wang2025generalizable, wang2025path} the core philosophy of leveraging physics-based structure decomposition and alignment to achieve environment generalization. However, the learning task of CSI feedback in this paper differs fundamentally. \cite{wang2025generalizable} addresses channel extrapolation, a supervised learning task where the goal is prediction accuracy, and employs path-level modeling. \cite{wang2025path} proposes a path evolution model for self-sustaining channel digital twins, focusing on temporal path dynamics. In contrast, this work tackles unsupervised CSI compression/reconstruction under feedback overhead constraints. This necessitates our novel designs of a cluster-based distribution shift model, an SVD-based real-time decoupling algorithm, a hybrid cluster-number estimator for robustness, and the EG-CsiNet structure with a metadata feedback scheme. The major contributions of this work can be summarized below: 
\begin{itemize}
    \item We propose a distribution shift model of the cluster-based channel as a foundation to understand the environment-generalizability, which comprises the distribution shift of cluster number, multi-cluster dependency, and single-cluster response. 
    \item We design a physics-based distribution alignment approach comprising multi-cluster decoupling and fine-grained alignment, which can effectively address the cross-environment distribution shift of the channel.
    \item We implement practical algorithms in the physics-based distribution alignment. Specifically, an efficient SVD-based multi-cluster decoupling algorithm is proposed based on the EYM theorem, which avoids the path-level parameter estimation and can support real-time CSI feedback. Additionally, a robust cluster number estimation is designed against the downlink channel estimation error. 
    \item We propose EG-CsiNet as a universal learning framework with the physics-based distribution alignment. The training and inference of EG-CsiNet are designed to enhance generalization, where the feedback overhead and model parameter complexity are also analyzed.
\end{itemize}

\textit{Notations:} $\mathbb{R}^{m\times n}$ and $\mathbb{C}^{m\times n}$ denote the real and complex spaces with dimension $m\times n$ and $\rm{j}=\sqrt{-1}$; $(\cdot)^{T}$, $\text{conj}(\cdot)$, and $(\cdot)^{H}$ denote the transpose, conjugate, and Hermitian transpose, respectively; $\otimes$ and $\odot$ stand for the Kronecker product and Hadamard product; $\text{rank}(\mathbf{X})$ denotes the rank of matrix $\mathbf{X}$; $\Vert\mathbf{X}\Vert_{F}$ denotes the Frobenius norm of matrix $\mathbf{X}$ and $\Vert\mathbf{x}\Vert_{2}$ stands for Euclidean norm of vector $\mathbf{x}$; $[x]$ $\lfloor x \rfloor$, $\lceil x\rceil$ denote the round, floor, and ceiling of real scalar $x$; $\delta(x)$ denotes the Dirac function; $\mathbb{E}\{x\}$ denotes the statistical expectation of random variable $x$; $f\circ g$ denotes the composition of function $f$ and $g$. 

\section{Problem Formulation and Key Solution} 
Firstly, the problem of the generalization challenge in deep learning-based CSI feedback is formulated in Sec.~\ref{subsec: problem}. Next, the distribution shift model for the cluster-based channel is proposed in Sec.~\ref{subsec: distribution shift model}. Then, the physics-based distribution alignment is proposed in Sec.~\ref{subsec: address shift} as a key solution to address the generalization challenge. 
\subsection{Problem Formulation: Generalization Challenge of Deep Learning-Based CSI Feedback} 
\label{subsec: problem}
Consider an FDD massive MIMO system with $N_{\text{T}}$ antennas and bandwidth $B$, which serves a single-antenna UE. After downlink channel estimation, the estimated CSI matrix at the user can be formulated as $\mathbf{H}=[\mathbf{h}_{1},\mathbf{h}_{2},\ldots,\mathbf{h}_{N_{\rm{C}}}]\in\mathbb{C}^{N_{\rm T}\times N_{\rm C}}$, where $\mathbf{h}_{k}\in\mathbb{C}^{N_{\rm{T}}\times 1}$ denotes the CSI of $k$th subcarrier and $N_{\rm{C}}$ denotes the number of subcarriers. Firstly, the CSI matrix $\mathbf{H}$ is transformed into the angular-delay domain $\widetilde{\mathbf{H}}=\mathbf{F}_{\rm{a}}\mathbf{H}\mathbf{F}_{\rm{d}}^{H}$ at the UE side, where $\mathbf{F}_{\rm a}\in\mathbb{C}^{N_{\rm{T}}\times N_{\rm{T}}}$ and $\mathbf{F}_{\rm{d}}\in\mathbb{C}^{N_{\rm{C}}\times N_{\rm{C}}}$ denote angular-domain representation matrix and the normalized discrete Fourier transformation (DFT) matrix. Compared to the original CSI matrix $\mathbf{H}$, the transformed $\widetilde{\mathbf{H}}$ exhibits obvious sparsity in the angular-delay domain, which facilitates low-overhead feedback. Then, $\widetilde{\mathbf{H}}$ is compressed by the encoder NN $f_{\text{en}}(\cdot)$ to generate a low-dimensional codeword $\mathbf{c}=f_{\rm{en}}(\widetilde{\mathbf{H}})\in\mathbb{R}^{M\times 1}$, where $M\ll N_{\rm{T}}N_{\rm{C}}$ denotes the codeword dimension. Next, the compressed codeword $\mathbf{c}$ is quantized into bits $\mathbf{b}=Q(\mathbf{c})$, where $Q(\cdot)$ denotes the quantization operation. At the BS side, the received quantized bits $\mathbf{b}$ are input into the decoder NN $f_{\text{de}}(\cdot)$ to generate the reconstructed channel. To optimize the learnable parameters in the NNs of the encoder and decoder, the mean square error (MSE) loss function 
\begin{equation}
    \label{equ: mse loss}
    \mathcal{L}=\Vert \widetilde{\mathbf{H}}-f_{\rm{de}}(Q(f_{\rm{en}}(\widetilde{\mathbf{H}})))\Vert_{F}^{2}
\end{equation}
is adopted during the offline training phase. Assume the training dataset $\mathcal{D}^{(\rm c)}$ for the encoder/decoder NNs is drawn from a distribution $P^{(\rm c)}$. Then, the parameters of $f_{\rm en}$ and $f_{\rm de}$ are optimized by minimizing $\mathcal{L}$ over $\mathcal{D}^{(\rm c)}$. Thus, the trained encoder and decoder can effectively compress/decompress the CSI samples in the distribution $P^{(\rm c)}$. However, the compression/decompression capability of the trained encoder/decoder cannot be guaranteed for the OOD channel samples. In the practical deployment in diverse wireless environments, the distribution of CSI samples is environment-dependent, where a large amount of OOD channel samples is inevitable in the new environments. Thus, the deep learning-based CSI feedback is faced with the generalization challenge, where the performance severely degrades in new environments.

\subsection{Problem Analysis: Distribution Shift Model for Cluster-Based Channel}
\label{subsec: distribution shift model}
The modeling of the cross-environment distribution shift of the channel samples is vital to understanding and enhancing model generalizability. To characterize the propagation of paths in practical wireless environments, the cluster-based massive MIMO channel model \cite{3gpp.38.901,tcom_ma_2024_deep} is adopted. Explicitly, the channel is composed of $N_{\rm cl}$ clusters, where the $l$th cluster contains $N_{{\rm p}, l}$ physical paths exhibiting similar angle and delay parameters. Without loss of generality, we assume a uniform antenna array (UPA) is equipped at the BS, where the numbers of horizontal and vertical antennas are set as $N_{\rm h}$ and $N_{\rm v}$, respectively. Then, the channel $\mathbf{h}_{k}$ of $k$th subcarrier can be modeled as 
\begin{equation}
    \label{equ: channel model}
    \mathbf{h}_{k}=\sum_{l=1}^{N_{\rm cl}}\sum_{i=1}^{N_{{\rm p}, l}}\alpha_{l,i}e^{-{\rm{j}}2\pi k\Delta f\tau_{l,i}}\mathbf{a}(\phi_{l,i},\theta_{l,i}),
\end{equation}
where $\Delta f=\frac{B}{N_{\rm C}}$ denotes the subcarrier spacing; $\alpha_{l,i},\phi_{l,i},\theta_{l,i},\tau_{l,i}$ stand for the complex gain, azimuth angle of departure (AAoD), elevation angle of departure (EAoD), and delay of the $i$th path in the $l$th cluster; $\mathbf{a}(\phi,\theta)=\mathbf{a}^{(\rm h)}(\phi,\theta)\otimes\mathbf{a}^{(\rm v)}(\theta)$ denotes the steering vector of half-wavelength antenna array, where \cite{jsac_yin_2020_Addressing}
\begin{equation}
    \label{equ: array response}
    \begin{aligned}
    \mathbf{a}^{(\rm h)}(\phi,\theta)&=\left[1, e^{{\rm j}\pi\sin{\phi}\sin{\theta}},\ldots,e^{{\rm j}(N_{\rm h}-1)\pi\sin{\phi}\sin{\theta}}\right]^{T},\\
    \mathbf{a}^{(\rm v)}(\theta)&=\left[1, e^{{\rm j}\pi\cos{\theta}},\ldots,e^{{\rm j}(N_{\rm v}-1)\pi\cos{\theta}}\right]^{T}.
    \end{aligned}
\end{equation}

Thus, the CSI matrix $\mathbf{H}$ can be further reformulated as 
\begin{equation}
    \label{equ: CSI matrix}
    \mathbf{H}=\sum_{l=1}^{N_{\rm cl}}\sum_{i=1}^{N_{{\rm p}, l}}\alpha_{l,i}\mathbf{a}(\phi_{l,i},\theta_{l,i})\mathbf{b}(\tau_{l,i})^{H}=\sum_{l=1}^{N_{\text{cl}}}\mathbf{H}_{l},
\end{equation}
where $\mathbf{H}_{l}=\sum_{i=1}^{N_{{\rm p},l}}\alpha_{l,i}\mathbf{a}(\phi_{l,i},\theta_{l,i})\mathbf{b}(\tau_{l,i})^{H}$ denotes the response of $l$th cluster and frequency domain response vector $\mathbf{b}(\tau)\in\mathbb{C}^{N_{\rm C}\times 1}$ can be represented by 
\begin{equation}
    \label{equ: freq resp}
    \mathbf{b}(\tau)=\left[1,e^{{\rm j}2\pi \Delta f\tau},\ldots,e^{{\rm j}2\pi(N_{\rm C}-1)\Delta f\tau}\right]^{T}.
\end{equation}

Based on the cluster-based channel model, the structure of the channel distribution shift is analyzed as follows. For the massive MIMO system in a specific environment, the electromagnetic wave interacts with the objects within the environment, where multiple clusters are yielded. Intuitively, the object density determines the distribution of the number of clusters. Owing to the unique geometrical layouts and the electromagnetic properties of the objects in the propagation channel, the path parameters of different clusters are not independent but exhibit complex dependencies. Hereby, the number of clusters and inter-cluster dependencies can be merged as multi-cluster structure. Additionally, the path parameters within a cluster are determined by the locations and materials of the interacted objects in the channel, whose distribution is also environment-dependent. Owing to the diverse user distributions and object layouts, both the distribution of multi-cluster structure and single-cluster response obviously vary across different environments, resulting in a significant distribution shift of the cluster-based channel. 

\begin{figure}[t]
    \vspace{-5pt}
        \centering
        \includegraphics[width=0.45\textwidth]{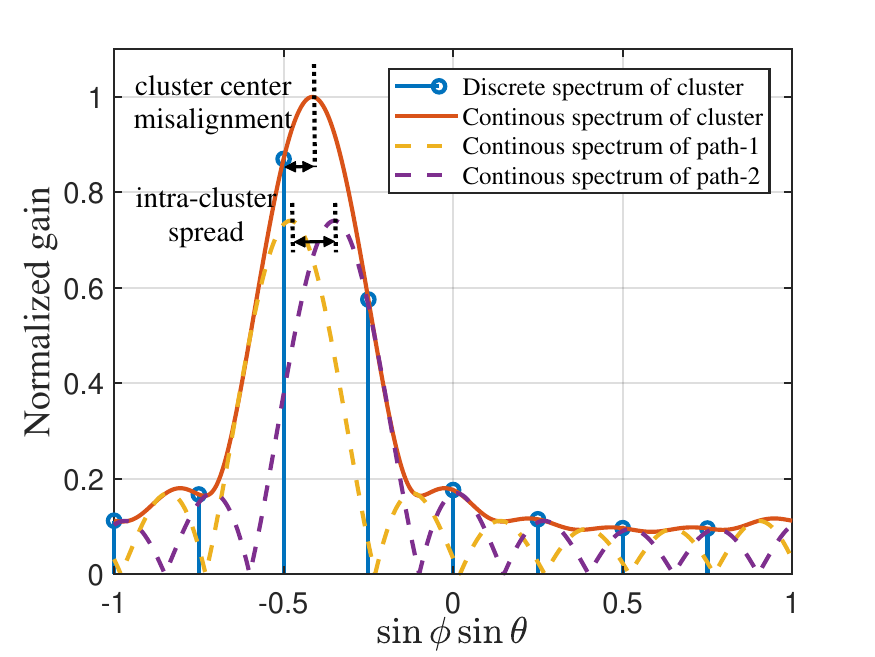}
    \captionsetup{font=footnotesize}
    \vspace{-5pt}
    \caption{Power leakage effect of a two-path cluster in the horizontal angular domain. The AoDs of the two paths are set as $\sin(\phi_{l,1})\sin(\theta_{l,1})=-0.48$ and $\sin(\phi_{l,2})\sin(\theta_{l,2})=-0.35$.}
    \label{fig: power_leakage}
    \vspace{-15pt}
\end{figure}

Further, the distribution shift of single-cluster response is investigated in the angular-delay domain. For BS equipped with UPA, the angular representation matrix can be represented as $\mathbf{F}_{\rm a}=\mathbf{F}_{\rm h}\otimes\mathbf{F}_{\rm v}$, where $\mathbf{\mathbf{F}}_{\rm h}\in\mathbb{C}^{N_{\rm h}\times N_{\rm h}}$ and $\mathbf{F}_{\rm v}\in\mathbb{C}^{N_{\rm v}\times N_{\rm v}}$ denote the normalized DFT matrices. Define the center AAoD, EAoD, and delay of $l$th cluster by $\phi_{l}$, $\theta_{l}$, and $\tau_{l}$. Then, the $(n,m)$th element in the angular-delay representation $\widetilde{\mathbf{H}}_{l}=\mathbf{F}_{\rm a}\mathbf{H}_{l}\mathbf{F}_{\rm d}^{H}$ of $l$th cluster can be represented by 
\begin{equation}
    \label{equ: element}
    \begin{aligned}
    &[\widetilde{\mathbf{H}}_{l}]_{n,m}\!=\!\sum_{i=1}^{N_{{\rm p},l}}\frac{\alpha_{l,i}}{\sqrt{N_{\rm T}N_{\rm C}}}\Big(\sum_{i_1=0}^{N_{\rm h}-1}e^{{\rm j}\pi i_{1}\big(\!\sin(\phi_{l,i})\sin(\theta_{l,i})\!-\!\frac{2 n_{1}}{N_{\rm h}}\big)}\Big)\times\\
    &\!\Big(\sum_{i_2=0}^{N_{\rm v}-1}e^{{\rm j}\pi i_{2}\big(\!\cos(\theta_{l,i})-\frac{2 n_{2}}{N_{\rm v}}\big)}\Big)\times\Big(\sum_{i_3=0}^{N_{\rm C}-1}e^{{\rm j}2\pi i_{3}\big(\frac{m}{N_{\rm C}}-\!\Delta f \tau_{l,i}\big)}\Big)\\
    &\!=\!\sum_{i=1}^{N_{{\rm p},l}}\alpha_{l,i}D_{N_{\rm{h}}}(k_{l}^{(\rm h)}\!+\!r_{l,i}^{(\rm h)}\!-\!n_1)\!\times\! D_{N_{\rm v}}(k_{l}^{(\rm v)}\!+\!r_{l,i}^{(\rm v)}\!-\!n_2)\\
    &\times D_{N_{\rm C}}(k_{l}^{(\rm d)}\!+\!r_{l,i}^{(\rm d)}\!-\!m),
    \end{aligned}
\end{equation}
where $D_{N}(x)=\frac{1}{\sqrt{N}}\sum_{n=0}^{N-1}e^{{\rm j} \frac{2\pi nx}{N}}=\frac{\sin(\pi x)}{\sqrt{N}\sin(\pi x/N)}e^{{\rm j}\left(\frac{N-1}{N}\right)\pi x}$, horizontal index $n_{1}=\lfloor n /N_{\rm v}\rfloor$, vertical index $n_2=n-n_1 N_{\rm v}$, horizontal angular-domain peak index $k_{l}^{(\rm h)}=\left[N_{\rm h}\sin(\phi_{l})\sin(\theta_{l})/2\right]$, vertical angular-domain peak index $k_{l}^{(\rm v)}=\left[N_{\rm v}\cos(\theta_{l})/2\right]$, delay-domain peak index $k_{l}^{(\rm d)}=\left[B \tau_{l}\right]$, and the residues $r_{l,i}^{(\rm h)}=N_{\rm h}\sin(\phi_{l,i})\sin(\theta_{l,i})/2-k_{l}^{(\rm h)}$, $r_{l,i}^{(\rm v)}=N_{\rm v}\cos(\theta_{l,i})/2-k_{l}^{(\rm v)}$, $r_{l,i}^{(\rm d)}=B\tau_{l,i}-k_{l}^{(\rm d)}$. When the residues of different paths within a cluster approach zero, elements in $\widetilde{\mathbf{H}}_{l}$ can be reformulated as
\begin{equation}
    \label{equ: zero residue}
    [\widetilde{\mathbf{H}}_{l}]_{n,m}=C\delta\left(n-N_{\rm v}k^{(\rm h)}_l-k_{l}^{(\rm v)}\right)\delta\left(m-k_{l}^{(\rm d)}\right),
\end{equation}
where constant $C=\sqrt{N_{\rm T}N_{\rm C}}\sum_{i=1}^{N_{{\rm p},l}}\alpha_{l,i}$. Then, the power of $\widetilde{\mathbf{H}}_{l}$ concentrates in a single grid. However, due to the limited number of antennas and subcarriers, residues $(r_{l,i}^{(\rm h)}, r_{l,i}^{(\rm v)}, r_{l,i}^{(\rm d)})$ are non-zeros and lead to the power leakage effect in the angular-delay domain \cite{tcom_ma_2021_deep}, which is explained as follows. Note that the roots of $D_N(x)$ are all integers. For non-zero residues $(r_{l,i}^{(\rm h)}, r_{l,i}^{(\rm v)}, r_{l,i}^{(\rm d)})$, $D_{N_{\rm{h}}}(k_{l}^{(\rm h)}\!+\!r_{l,i}^{(\rm h)}\!-\!n_1),D_{N_{\rm v}}(k_{l}^{(\rm v)}\!+\!r_{l,i}^{(\rm v)}\!-\!n_2),D_{N_{\rm C}}(k_{l}^{(\rm d)}\!+\!r_{l,i}^{(\rm d)}\!-\!m)$ in \eqref{equ: element} are non-zeros for integers $(n_1, n_2, m)$, which results in the power leakage effect. Then, the power leakage effect consequently leads to an increase in the normalized power contained in the off-peak elements of the angular-delay domain, i.e., those elements \((n', m')\) satisfying \((n',m') \neq \arg\max_{n,m} |[\widetilde{\mathbf{H}}]_{n,m}|^2\). Without loss of generalizability, the power leakage effect in the horizontal angular domain is presented as an example. Hereby, the residue $r^{(\rm h)}_{l,i}$ in the horizontal angular domain can be reformulated as 
\begin{equation}
    \label{equ: residue}
    \begin{aligned}
    r_{l,i}^{(\rm h)}&=\underbrace{\frac{N_{\rm T}\sin(\phi_{l})\sin(\theta_{l})}{2}-k_{l}^{(\rm h)}}_{\text{cluster center misalignment}}\\&+\underbrace{\frac{N_{\rm T}}
    {2}\big(\sin(\phi_{l,i})\sin(\theta_{l,i})-\sin(\phi_{l})\sin(\theta_{l})\big)}_{\text{intra-cluster spread}}.
    \end{aligned}
\end{equation}
Therefore, the power leakage effect of a cluster is governed by both cluster center misalignment and intra-cluster spread, where an example is also illustrated in Fig.~\ref{fig: power_leakage}. For a given cluster, the peak indexes and cluster center misalignment are determined by the interacted positions along the cluster, which depend on the user/BS positions and the geometrical object layouts in the environments. Meanwhile, the materials of the objects also affect the cluster peak complex gain and intra-cluster spreads. Thus, distributions of peak indexes, power leakage, and complex path gain of a cluster in the angular-delay domain drastically shift across environments, which leads to the distribution shift of a single cluster. 

\subsection{Key Solution: Physics-based Distribution Alignment}
\label{subsec: address shift}
As illustrated in the block (b) of Fig.~\ref{fig: framework}, the physics-based distribution alignment is introduced to address the CSI distribution shift in Sec.~\ref{subsec: distribution shift model}, where the two modules of multi-cluster decoupling and fine-grained alignment are discussed as follows. 

Multi-cluster decoupling module can address the distribution shift of the multi-cluster structure. Based on the channel model in \eqref{equ: CSI matrix}, the response of different clusters can be represented in a unified form, which supports the cluster-wise feedback. Therefore, the UE can first apply multi-cluster decoupling to the original channel, and the encoder and decoder modules can individually compress and reconstruct each decoupled cluster. Based on the cluster-wise feedback manner, the NNs in the encoder and decoder will not fit the distribution of the number and dependencies of the decoupled clusters, which intuitively addresses the distribution shift of the multi-cluster structure \cite{wang2025path,wang2025generalizable}. 

Further, fine-grained alignment is individually applied to the decoupled cluster components to address the single-cluster distribution shift. Specifically, the peak indexes of decoupled cluster components are precisely searched and aligned to a fixed position, and the distribution shifts of the power leakage effect and complex gain are also mitigated. With the multi-cluster decoupling and fine-grained alignment, the distribution shift of the multi-cluster CSI in Sec.~\ref{subsec: distribution shift model} can be effectively addressed, and the environment-generalizability can be greatly enhanced, which is interpretable in physics. 

\section{Implementation of Physics-Based Distribution Alignment}
\label{sec: address}
In Sec.~\ref{subsec: decoupling}, an efficient multi-cluster decoupling algorithm is proposed, which mitigates the distribution shift of the multi-cluster structure. Then, the fine-grained alignment is detailed in Sec.~\ref{subsec: alignment}, which addresses the distribution shift of single-cluster response. Further, robust estimation of the cluster number is proposed in Sec.~\ref{subsec: noise robustness} against downlink channel estimation error.

\subsection{Multi-Cluster Decoupling}
\label{subsec: decoupling}
The objective of multi-cluster decoupling is to decompose the original channel $\mathbf{H}$ into a summation form $\mathbf{H}\approx\sum_{l=1}^{\widehat{R}}\mathbf{C}_{l}$, where $\mathbf{C}_{l}\in\mathbb{C}^{N_{\rm T}\times N_{\rm C}}$ denotes the response of $l$th decoupled cluster, and $\widehat{R}$ denotes the number of decoupled cluster. Intuitively, the power distribution of each decoupled component is clustered in the angular-delay domain. In our earlier work \cite{wang2025generalizable}, a path extraction algorithm based on the space-alternating generalized expectation-maximization (SAGE) algorithm \cite{jsac_Fleury_1999_sage} and density-based spatial clustering of applications with noise (DBSCAN) algorithm \cite{tods_Schubert_2017_dbscan} is proposed. Explicitly, the intermediate path-level parameters are first estimated via the SAGE algorithm, and then the cluster-level responses are yielded via DBSCAN clustering. However, the complexity of SAGE-based intermediate parameter estimation is relatively high, encountering real-time deployment challenges in dynamic scenarios.

To facilitate real-time CSI feedback, a multi-cluster decoupling algorithm with low complexity is proposed. Specifically, different clusters are directly decoupled based on cluster-level properties, which can avoid the intermediate path-level parameter estimation and directly reduce computation complexity. To this end, the intra-cluster and inter-cluster properties are investigated as follows to support the efficient multi-cluster decoupling.

\begin{table}[t]
  \centering
  \belowrulesep=0.5pt
  \aboverulesep=0pt
  \begin{threeparttable}
  \captionsetup{font=footnotesize}
  \caption{Rank-one dominance of individual cluster in 3GPP 38.901 UMa scenario.}
    \label{tab: rank-one dominance}%
    \begin{tabular}{c|c|c}
    \toprule
    \multirow{2}[2]{*}{\centering scenarios} & \multicolumn{2}{c}{UMa}\\
\cmidrule{2-3}          & LOS   & NLOS \\
    \midrule
    Intra-cluster AoD spread & 5$^\circ$ & 2$^\circ$ \\
    \midrule
    Intra-cluster delay spread & 4.7 ns & 4.7 ns\\
    \midrule
    Concentration $\xi$ & 0.993 & 0.994\\
    \bottomrule
    \end{tabular}%
  \end{threeparttable}
  \vspace{-15pt}
\end{table}%

\subsubsection{Cluster-Level Properties}
Due to the clustering nature, the paths within a cluster exhibit similar AoD and delay parameters. When the intra-cluster spread is smaller than system resolutions, the steering vector $\mathbf{a}(\phi)$ and frequency domain response vector $\mathbf{b}(\tau)$ of the paths within a cluster are highly linearly dependent. Thus, the rank of cluster response $\mathbf{H}_{l}$ is approximately one. Consider a numerical example for further justification. To quantify the rank-one dominance of $\mathbf{H}_{l}$, the concentration $\xi=\sigma_{1}^2/\Vert\mathbf{H}_{l}\Vert_{F}^2$ can be defined, where $\sigma_{1}$ denotes the largest singular value of $\mathbf{H}_{l}$. Consider a BS with 8 horizontal antennas, where the bandwidth is 10 MHz and the number of subcarriers is 32. Then, based on the intra-cluster spread and offset specifications of the urban macro (UMa) scenario in 3GPP 38.901 document \cite{3gpp.38.901}, power portion $\xi$ in the line of sight (LOS) and non-line of sight (NLOS) status are shown in Table~\ref{tab: rank-one dominance}. It can be found that the power portion $\xi>0.99$ is held in different conditions, which verifies the rank-one approximation for a single cluster response. Thus, the intra-cluster property of rank-one dominance is given as follows.

\begin{proposition}
\label{prop: rank-1}
For a single cluster, $\text{rank}(\mathbf{H}_{l})\approx1$ can be approximated when the intra-cluster spread is smaller than system resolution. 
\end{proposition}

\begin{figure}[t]
    \vspace{-10pt}
    \captionsetup{font=footnotesize,justification=centering}
    \subfloat[LOS scenario]{\includegraphics[width=0.24\textwidth]{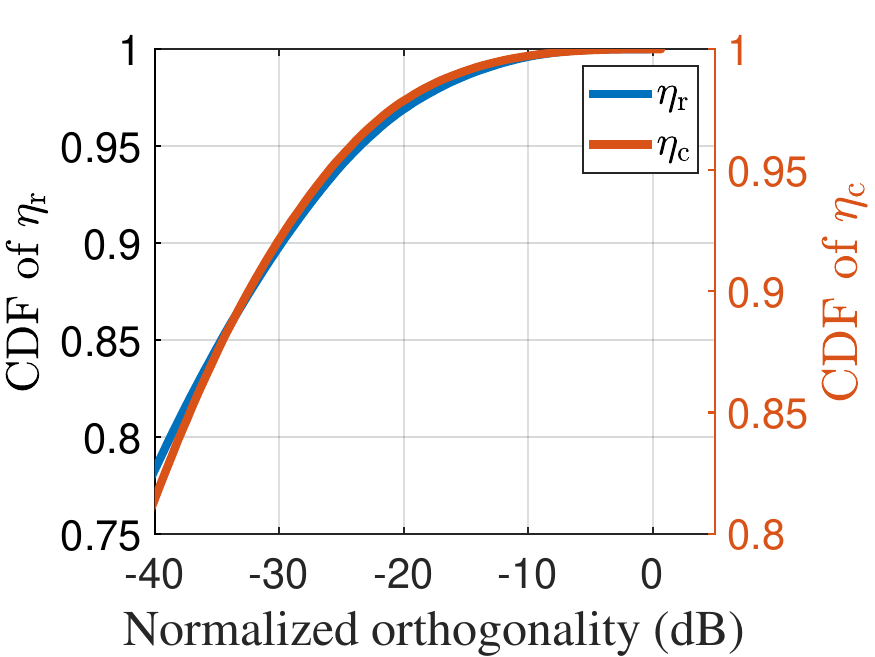}}\captionsetup{font=footnotesize,justification=centering}
    \subfloat[NLOS scenario]{\includegraphics[width=0.24\textwidth]{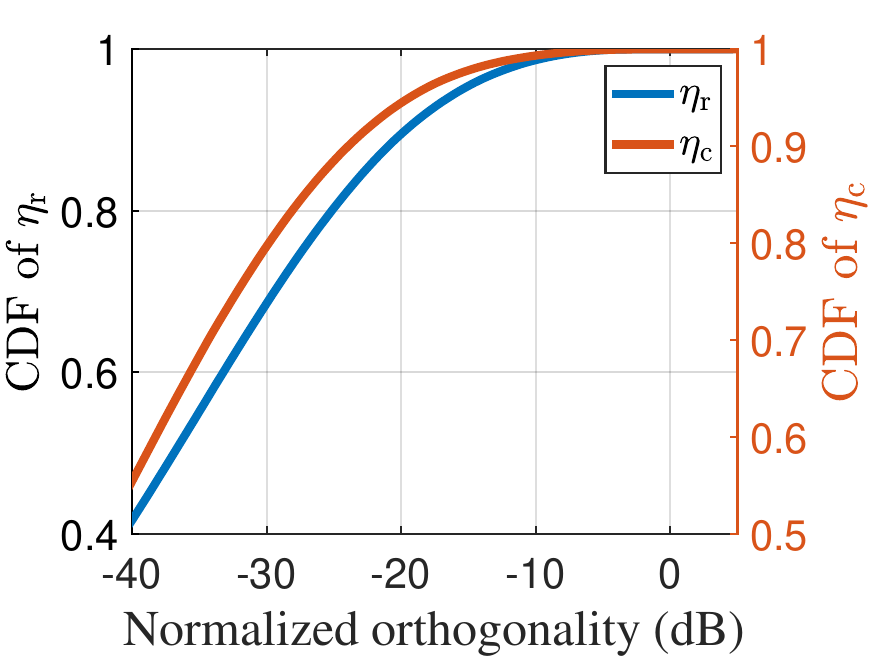}}\captionsetup{font=footnotesize,justification=raggedright}
    \captionsetup{font=small}
    \caption{Numerical validation of cluster orthogonality.}
    \label{fig: orthogonality}
    \vspace{-10pt}
\end{figure}

Based on the real-world channel measurement campaign, the power of the channel is concentrated in the LOS path and single-hop cluster \cite{tvt_liu_2024_shared}. According to the geometrical relationship, the AoD and delay parameters of the cluster are determined by the location of the interacted scatterer. In the usual scenarios, the deployments of scatterers are asymmetrical to the BS and UE. Hence, the spatial separation and asymmetrical deployment of the scatterers result in both distinct AoD and delay parameters between different clusters. As a result, orthogonality is held for the steering vectors of the paths from different clusters, i.e., $\mathbf{a}^{H}(\phi_{l,i},\theta_{l,i})\mathbf{a}(\phi_{l^{\prime},i^{\prime}},\theta_{l^{\prime},i^{\prime}})\approx0$. Similarly, the frequency domain response vectors of different clusters are orthogonal as well. Additionally, a numerical experiment based on 3GPP 38.901 UMa channel model \cite{3gpp.38.901} is provided to support the orthogonality assumption. For the clusters $\mathbf{H}_{l}$ and $\mathbf{H}_{l^\prime}$ in channel $\mathbf{H}$, the normalized row/column orthogonality $\eta_{\rm r}/\eta_{\rm c}$ can be defined, which are calculated by
\begin{equation}
    \eta_{\rm r}=\frac{\Vert\mathbf{H}_{l}^{H}\mathbf{H}_{l^\prime}\Vert_{F}}{\Vert\mathbf{H}^{H}\mathbf{H}\Vert_{F}},~~\eta_{\rm c}=\frac{\Vert\mathbf{H}_{l}\mathbf{H}_{l^\prime}^{H}\Vert_{F}}{\Vert\mathbf{H}\mathbf{H}^H\Vert_{F}}.
\end{equation}
Intuitively, a small $\eta_{\rm r}$ and $\eta_{\rm c}$ can indicate orthogonality of different clusters is approximately held in the channel. Then, the cumulative density functions (CDFs) of $\eta_{\rm r}$ and $\eta_{\rm c}$ in LOS and NLOS scenarios are plotted in Fig.~\ref{fig: orthogonality}. It can be found that 90th percentile of $\eta_{\rm r}$ and $\eta_{\rm c}$ can achieve -30$\sim$-20 dB in both LOS and NLOS scenarios, which indicates the approximate orthogonality among different clusters. Therefore, both the rowspace and columnspace of different cluster components are orthogonal, where the dual-orthogonality is formulated below.

\begin{proposition}
\label{prop: orthogonality}
For two different clusters $\mathbf{H}_{l}$ and $\mathbf{H}_{l^{\prime}}$, $\mathbf{H}_{l}^{H}\mathbf{H}_{l^{\prime}}\approx\mathbf{0}$ and $\mathbf{H}_{l}\mathbf{H}_{l^{\prime}}^{H}\approx\mathbf{0}$ can be approximated, especially when the scatterers exhibit large spatial separation. 
\end{proposition}
\subsubsection{SVD-Based Multi-Cluster Decoupling}
Based on intra-cluster property in \textbf{Proposition}~\ref{prop: rank-1} and inter-cluster property in \textbf{Proposition} \ref{prop: orthogonality}, multi-cluster decoupling is formulated as follows. Explicitly, decoupled clusters $\{\mathbf{C}_{l}\}_{l=1}^{\widehat{R}}$ can be optimized by 
\begin{subequations}
\label{equ: cluster optimization}
\begin{align}
    \min_{\{\mathbf{C}_{l}\}_{l=1}^{\widehat{R}}} \quad & \Vert\mathbf{H}-\sum_{l=1}^{\widehat{R}}\mathbf{C}_{l}\Vert_{F}, \label{subeq: obj}\\
    \text{s.t.} \quad & \mathcal{C}_1: \text{rank}(\mathbf{C}_{l})=1, \quad 1\leq l\leq\widehat{R} \label{subeq: C1}\\
    & \mathcal{C}_2: \mathbf{C}_{l}^{H}\mathbf{C}_{l^{\prime}}=\mathbf{0},\quad\mathbf{C}_{l}\mathbf{C}_{l^{\prime}}^{H}=\mathbf{0}, \quad \forall l\neq l^{\prime} \label{subeq: C2},
\end{align}
\end{subequations}
where the constraints $\mathcal{C}_{1}$ and $\mathcal{C}_{2}$ are derived from \textbf{Proposition}~\ref{prop: rank-1} and \ref{prop: orthogonality}, respectively \footnote{Note that the optimization variable $\mathbf{C}_{l}$ is distinct from the physical cluster response $\mathbf{H}_{l}$. Due to the limited number of antennas and bandwidth, the ground-truth $\mathbf{H}_{l}$ cannot be resolved from the channel observation $\mathbf{H}$. To this end, we aim to approximate $\mathbf{H}_{l}$ via the multi-cluster decoupling optimization in \eqref{equ: cluster optimization}. Practically, owing to approximation in hard-constraints, i.e., $\mathcal{C}_1$ and $\mathcal{C}_2$, the decoupled cluster $\mathbf{C}_{l}$ cannot perfectly represent the physical cluster $\mathbf{H}_{l}$.}. Based on the EYM theorem in low-rank matrix approximation \cite{eckart1936approximation,mirsky1960symmetric}, the closed-form solution of \eqref{equ: cluster optimization} can be derived as follows. 
\begin{theorem}
\label{theo: optimal cluster}
Let the SVD of $\mathbf{H}$ be given by $\mathbf{H}=\sum_{l=1}^{\text{rank}(\mathbf{H})}\sigma_{l}\mathbf{u}_{l}\mathbf{v}_{l}^{H}$, where $\sigma_{l}$ denotes the $l$th largest singular value of $\mathbf{H}$, $\mathbf{u}_{l}$ and $\mathbf{v}_{l}$ are the singular vectors. Then, the optimal solution of \eqref{equ: cluster optimization} can be represented by $\mathbf{C}_{l}^{\star}=\sigma_{l}\mathbf{u}_{l}\mathbf{v}_{l}^{H}$ for $1\leq l\leq\widehat{R}$. 
\end{theorem}
\begin{IEEEproof}
Details are presented in Appendix~\ref{appdix: theo 1 proof}.
\end{IEEEproof}
Compared to the conventional path extraction based on the SAGE algorithm and DBSCAN clustering, the proposed SVD-based multi-cluster decoupling avoids the intermediate path-level parameter estimation, and the calculation can be sped up via parallel computation methods \cite{math_algorithm_Feng_2024}, which facilitates the real-time application. Here, the number of decoupled clusters $\widehat{R}$ serves as an initial parameter in \eqref{equ: cluster optimization}, which is determined in Sec.~\ref{subsec: noise robustness}. 

\begin{remark}
    \label{remk: compare eigenvalue feedback}
    Proposed SVD-based multi-cluster decoupling is distinct from the eigenvector-based CSI feedback with multi-antenna UE \cite{wcl_evcsinet_2021_liu}. The eigenvector-based CSI feedback is originally designed to directly maximize the downlink beamforming gain \cite{twc_Ayach_2014_spatially}, instead of capturing the cluster-level features. The channel eigenvector $\mathbf{v}_{k}$ at $k$th subcarrier is obatained by solving $\lambda_k \mathbf{v}_k=(\mathbf{h}_k\mathbf{h}^H_k)\mathbf{v}_{k}$, where $\lambda_k$ denotes the largest eigenvalue of matrix \(\mathbf{h}_k\mathbf{h}^H_k\). Thus, the decomposition dimension is independent of the frequency domain and frequency/delay cluster features are not extracted in the channel eigenvectors $\{\mathbf{v}_{k}\}.$
\end{remark}

\subsection{Fine-Grained Alignment} 
\label{subsec: alignment}
The fine-grained alignment module aims to remove the bias of each decoupled cluster, which achieves the goal of distribution alignment. Here, the fine-grained alignment of an individual decoupled cluster $\mathbf{C}\in\mathbb{C}^{N_{\rm T}\times N_{\rm C}}$ is presented as an example. Based on the distribution model in Sec.~\ref{subsec: distribution shift model}, the distribution shift of cluster centers simultaneously results in the distribution shift of the peak indexes and the cluster center misalignment in the angular-delay domain. Motivated by our earlier work \cite{wang2025generalizable}, the oversampled Kronecker codebook \cite{tcom_ma_2024_deep}, and DFT codebook are employed to scan the fine-grained peak positions in the angular and delay domains, respectively. Denote the horizontal and vertical oversampling factors as $O_{\rm h}$ and $O_{\rm v}$. Then, the $(n_1,n_2)$th codeword $\mathbf{w}_{n_{1},n_{2}}^{(\rm a)}\in\mathbb{C}^{N_{\rm T}\times1}$ can be calculated by $\mathbf{w}_{n_{1},n_{2}}^{(\rm a)}=\mathbf{w}_{n_1}^{(\rm a,h)}\otimes\mathbf{w}_{n_2}^{(\rm a,v)}$, where
\begin{equation}
    \label{equ: angular codeword}
    \begin{aligned}
    \mathbf{w}_{n_1}^{(\rm a,h)}&=\left[1,e^{{\rm j}2\pi \frac{n_1}{O_{\rm h}N_{\rm h}}},\ldots,e^{{\rm j}2\pi \frac{n_1(N_{\rm h}-1)}{O_{\rm h}N_{\rm h}}}\right]^{T},\\
    \mathbf{w}_{n_2}^{(\rm a,v)}&=\left[1,e^{{\rm j}2\pi \frac{n_2}{O_{\rm v}N_{\rm v}}},\ldots,e^{{\rm j}2\pi \frac{n_2(N_{\rm v}-1)}{O_{\rm v}N_{\rm v}}}\right]^{T},
    \end{aligned}
\end{equation}
$0\leq n_1\leq O_{\rm h}N_{\rm h}-1$, and $0\leq n_2\leq O_{\rm v}N_{\rm v}-1$. Then, the fine-grained peak position in the angular domain can be calculated by 
\begin{equation}
    \label{equ: peak position angular}
    (n^\star_1,n_2^\star)=\mathop{\arg\max}_{n_1,n_2}\left\{\Vert(\mathbf{w}_{n_1,n_2}^{(\rm a)})^{H}\mathbf{C}\Vert_{2}^{2}\right\}.
\end{equation}
Similarly, the $m$th codeword ($0\leq m\leq N_{\rm C}-1$) for the delay-domain $O_{\rm d}$-oversampled DFT codebook can be formulated as
\begin{equation}
    \label{equ: delay codeword}
    \mathbf{w}_{m}^{(\rm d)}=\left[1,e^{{\rm j}2\pi \frac{m}{O_{\rm d}N_{\rm C}}},\ldots,e^{{\rm j}2\pi \frac{m(N_{\rm C}-1)}{O_{\rm d}N_{\rm C}}}\right]^{T},
\end{equation}
Then, the fine-grained delay-domain peak position is yielded by
\begin{equation}
    \label{equ: peak position delay}
    m^\star=\mathop{\arg\max}_{m}\left\{\Vert\mathbf{C}\mathbf{w}_{m}^{(\rm d)}\Vert_{2}^{2}\right\}. 
\end{equation}
Based on the property of DFT transformation, element-wise phase adjustment can be applied in $\mathbf{C}$ to align the angular-delay domain peak to a fixed position, where the phase adjustment matrix $\mathbf{S}\in\mathbb{C}^{N_{\rm T}\times N_{\rm C}}$ can be calculated by 
\begin{equation}
    \label{equ: phase adjust}
    \mathbf{S}=\text{conj}(\mathbf{w}^{(\rm a)}_{n^\star_1,n_2^\star})\otimes(\mathbf{w}_{m^\star}^{(\rm d)})^{T}.
\end{equation}
The proof of peak position alignment with matrix $\mathbf{S}$ is provided in Appendix~\ref{appdix: phase adjustment proof}. Based on the scanned fine-grained positions $(n^\star_1,n_2^\star,m^\star)$, the peak value of $\mathbf{C}$ can be calculated by $p=(\mathbf{w}_{n^\star_1,n^\star_2}^{(\rm a)})^{H}\mathbf{C}\mathbf{w}_{m^\star}^{(\rm d)}$. With the multi-cluster decoupling step, the peak value $p$ can reflect the complex gains of the paths within the cluster. Therefore, we can quantize the phase of peak value $p$ with $Q_{\rm p}$-bit uniform quantization. Explicitly, the $t$th codeword $\beta_{t}$ in the $Q_{\rm p}$-bit uniform phase quantization codebook can be calculated by
\begin{equation}
    \label{equ: phase codeword}
    \beta_{t}=\frac{2\pi t}{2^{Q_{\rm p}}},\quad t=0,\ldots,2^{Q_{\rm p}}-1.
\end{equation}
Then, peak phase $\angle p$ can be quantized with the codebook $\{\beta_{t}\}_{t=0}^{2^{Q_{\rm p}}-1}$ and the index $t^\star$ of the quantized phase $\beta_{t^\star}$ is yielded by 
\begin{equation}
    \label{equ: phase quantization}
    t^\star=\mathop{\arg\min}_{t}|\angle{p}-\beta_{t}|
\end{equation}
By applying a phase shift $-\beta_{t^\star}$ to each element in $\mathbf{C}$, the distribution shift of path gains within a cluster can also be mitigated. Based on the aforementioned processing, the aligned cluster component $\widetilde{\mathbf{C}}^{(\rm aln)}$ in the angular-delay domain is yielded by
\begin{equation}
    \label{equ: align}
    \widetilde{\mathbf{C}}^{(\rm aln)}=\mathbf{F}_{\rm a}(e^{-{\rm j}\beta_{t^\star}}\mathbf{S}\odot\mathbf{C})\mathbf{F}_{\rm d}^{H},
\end{equation}
which can effectively address the bias of each decoupled cluster, including peak indexes, cluster center misalignment, and complex gain. 

\subsection{Robust Estimation of Cluster Number}
\label{subsec: noise robustness}
In the aforementioned design, the clean channel matrix $\mathbf{H}$ without estimation error is assumed. Practically, the downlink channel is estimated from the received pilot symbols, which are affected by the additive noise at the UE. Based on the widely-used least square (LS) downlink channel estimation procedure with orthogonal pilots \cite{tvt_virtual_2020_zhao}, the estimated CSI matrix $\mathbf{H}^{(\rm e)}\in\mathbb{C}^{N_{\rm T}\times N_{\rm C}}$ can be formulated as $\mathbf{H}^{(\rm e)}=\mathbf{H}+\mathbf{N}^{(\rm e)}$, where $\mathbf{N}^{(\rm e)}$ denotes the estimation error. Thus, noise-robust design is essential for EG-CsiNet to guarantee feedback performance of different users in the environment.

Notably, the proposed SVD-based multi-cluster decoupling in EG-CsiNet is noise-robust, which is still applicable in a low-SNR regime. According to \textbf{Theorem}~\ref{theo: optimal cluster}, when the SVD-based multi-cluster decoupling is applied to $\mathbf{H}^{\rm (e)}$, the summation of decoupled clusters is a truncated SVD version of the $\mathbf{H}^{\rm (e)}$, which serves as a near-optimal approximation with the noised measurement \cite{tit_optshrink_2014_Nadakuditi}. Thus, SVD-based multi-cluster decoupling exhibits robust denoising capability, which can guarantee the performance of EG-CsiNet in the presence of channel estimation error.

\begin{figure*}[t]
        \centering
        \includegraphics[width=1\textwidth]{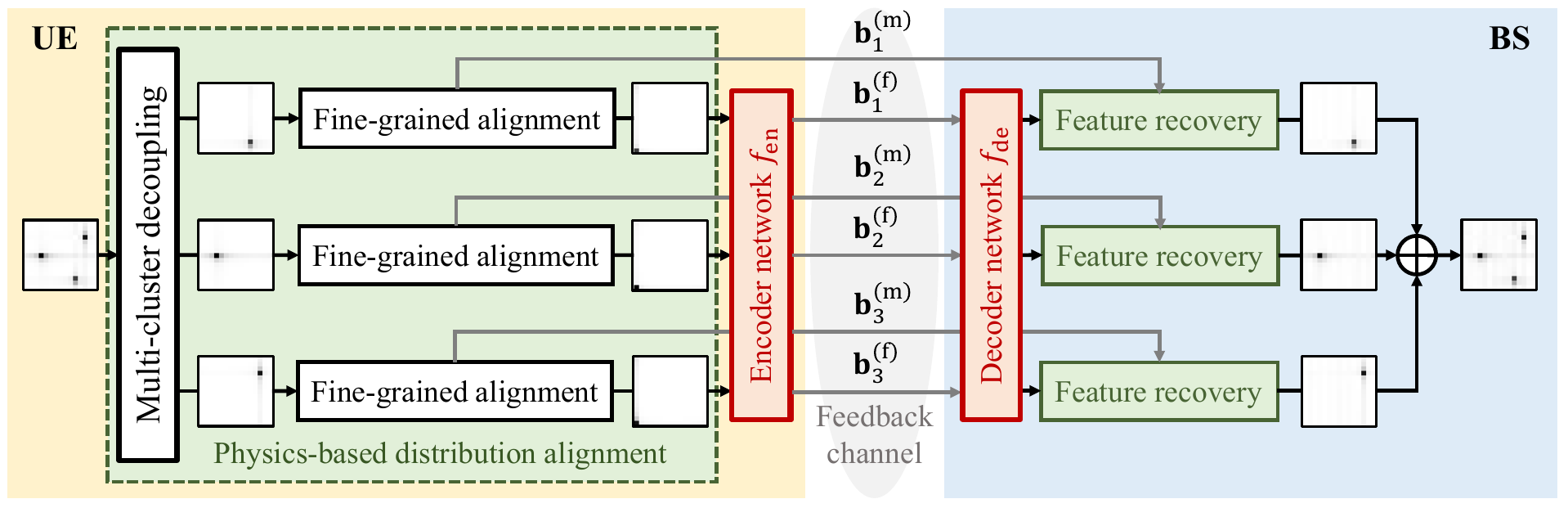}
    \captionsetup{font=footnotesize}
    \caption{Detailed structure of proposed EG-CsiNet, where three cluster components are decoupled as an example. The amplitudes of the angular-delay representation for input, output, and intermediate cluster components are also illustrated.}
    \label{fig: model}
    \vspace{-15pt}
\end{figure*}

To facilitate multi-cluster decoupling and efficient feedback in different SNR ranges, the number of decoupled clusters $\widehat{R}$ needs to be appropriately estimated from $\mathbf{H}^{(\rm e)}$. Explicitly, the clusters corrupted by estimation noise or weak clusters should be abandoned in order to minimize the feedback overhead while guaranteeing the feedback precision. To this end, a hybrid criterion based on the minimum description length (MDL) criterion \cite{tsp_rokota_2017_rank} and clip-threshold is proposed. The MDL criterion can effectively identify the number of components from the noised observation, which is derived from the perspective of information theory. Explicitly, denote the SVD of $\mathbf{H}^{(\rm e)}$ by $\mathbf{H}^{(\rm e)}=\sum_{i=1}^{\text{rank}(\mathbf{H}^{(\rm e)})}\widehat{\sigma}_{i}\widehat{\mathbf{u}}_{i}\widehat{\mathbf{v}}^{H}_{i}$, where $\widehat{\sigma}_{i}$ denotes the $i$th largest singular value of $\mathbf{H}^{(\rm e)}$. For MDL criterion, the number of decoupled clusters $\widehat{R}_{1}$ from $\mathbf{H}^{(\rm e)}$ is calculated by \footnote{Let $\mathbf{H}$ be a matrix of rank $r$ parameterized by $\Theta$. The MDL estimator $\widehat{R}_1$ in \eqref{equ: MDL} is originally defined as: $\widehat{R}_1 = \mathop{\arg\min}_{r} \left\{ -\log f(\mathbf{H} \mid \widehat{\Theta}^{(r)}) + k \log(N_{\text{C}}) \right\}$, where $f(\mathbf{H} \mid \widehat{\Theta}^{(r)})$ is the likelihood function, $\widehat{\Theta}^{(r)}$ is the maximum likelihood estimate of $\Theta$, and $k$ denotes the number of free parameters in $\Theta$. The complete derivation of \eqref{equ: MDL} is provided in \cite{tsp_detection_1985_wax}.}
\begin{equation}
    \begin{aligned}
    \label{equ: MDL}
    \widehat{R}_{1}\!&=\!\mathop{\arg\!\min}_{r}\Bigg\{\!-\!2N_{\rm C}(N_{\rm T}\!-\!r)\log\!\left(\frac{\prod_{i=r+1}^{N_{\rm T}}\widehat{\sigma}_{i}^{2/(N_{\rm T}-r)}}{\frac{1}{N_{\rm T}-r}\sum_{i=r+1}^{N_{\rm T}}\widehat{\sigma}_{i}^2}\right)\\
    &+r(2N_{\rm T}-r)\log(N_{\rm C})\Bigg\}.
    \end{aligned}
\end{equation}
Then, the estimation error in $\mathbf{H}^{(\rm e)}$ can be largely filtered out by truncating the largest $\widehat{R}_{1}$ components in SVD. In a relatively high-SNR regime, weak cluster components with a small power can also be detected based on the MDL criterion, which results in a relatively large $\widehat{R}_{1}$. Thus, to balance the CSI feedback precision and overhead in the high-SNR regime, the weak cluster components can be clipped by a pre-defined threshold $\eta\leq1$, where the number of decoupled clusters $\widehat{R}_{2}$ can be determined by 
\begin{equation}
    \label{equ: threshold}
    \widehat{R}_{2}=\min r, \quad \textnormal{s.t.}\sum_{i=1}^{r}\widehat{\sigma}_{i}^2\geq\eta\Vert\mathbf{H}^{(\rm e)}\Vert_{F}^{2}.
\end{equation}
The pre-defined threshold $\eta$ is determined by the required channel feedback accuracy. The selection of threshold $\eta$ should guarantee that the impact of the truncated SVD is negligible to the overall reconstruction precision. Explicitly, define the metric normalized missed detection error (NMDE) to quantify the effects of truncated components \cite{wang2025generalizable}, which is calculated by  
\begin{equation}
    \label{equ: NMDE}
    \text{NMDE}=\frac{\Vert\mathbf{H}-\sum_{l=1}^{\widehat{R}}\mathbf{C}_{l}^{\star}\Vert_{F}^2}{\Vert\mathbf{H}\Vert_{F}^2}.
\end{equation}
Thus, when the average $\text{NMDE}$ is far less than the target normalized mean square error (NMSE) of the reconstructed channel, the SVD truncation slightly impacts the overall channel reconstruction precision. Based on \eqref{equ: MDL} and \eqref{equ: threshold}, the number of decoupled clusters can be calculated by
\begin{equation}
    \label{equ: hat R final}
    \widehat{R}=\min\{\widehat{R}_{1},\widehat{R}_{2}\},
\end{equation}
which can robustly balance the feedback overhead and precision in different SNR ranges.

\section{EG-CsiNet: Generalizable CSI Feedback with Physics-Based Distribution Alignment}
In this section, the EG-CsiNet for CSI feedback is proposed. Firstly, the model training and inference in EG-CsiNet are presented in Sec.~\ref{subsec: training} to facilitate model generalizability. Next, the feedback overhead of EG-CsiNet is analyzed in Sec.~\ref{subsec: feedback overhead}. Then, the complexity of model parameters in EG-CsiNet is discussed in Sec.~\ref{subsec: params complexity}.

\label{sec: multi-cluster dcoupling}

\subsection{Model Training and Inference}
\label{subsec: training}
With the physics-based distribution alignment, the structure of the proposed EG-CsiNet is illustrated in Fig.~\ref{fig: model}. In the offline model training phase, the encoder and decoder NNs of EG-CsiNet are trained with the aligned clusters to address the CSI distribution shift. To this end, a training dataset of $\widetilde{\mathbf{C}}^{\text{(aln)}}$ should first be yielded based on the multi-cluster decoupling and fine-grained alignment steps. Then, the training loss of EG-CsiNet is defined as 
\begin{equation}
    \label{equ: aligned loss}
    \mathcal{L}_{\text{EG-CsiNet}}=\Vert\widetilde{\mathbf{C}}^{(\rm aln)}-f_{\rm de}(Q(f_{\rm en}(\widetilde{\mathbf{C}}^{(\rm aln)})))\Vert_{F}^{2}.
\end{equation}
Thus, the optimal NN parameters of $f_{\rm en}(\cdot)$ and $f_{\rm de}(\cdot)$ are yielded by minimizing $\mathcal{L}_{\text{EG-CsiNet}}$, which are retained in the online model inference phase after deployment. By applying the loss function $\mathcal{L}_{\text{EG-CsiNet}}$, the NNs of the encoder and decoder effectively compress and decompress the aligned cluster component $\widetilde{\mathbf{C}}^{(\rm aln)}$ in the distribution of the training dataset. Based on the analysis in Sec.~\ref{subsec: address shift}, the distribution of $\widetilde{\mathbf{C}}^{(\rm aln)}$ in the training and the unseen test datasets can be effectively aligned. Thus, the trained encoder and decoder can robustly compress and decompress the aligned cluster component $\widetilde{\mathbf{C}}^{(\rm aln)}$ in the unseen test environments. 

In the online model inference phase, the reconstructed CSI matrix that comprises multiple clusters is yielded, where the end-to-end procedure is illustrated in Fig.~\ref{fig: model}. At the UE side, multi-cluster decoupling and fine-grained alignment are applied to the $\mathbf{H}$, which yields the aligned cluster components and the related metadata $(n^\star_1,n^\star_2,m^{\star},t^\star)$. After the encoder NN, the compressed codeword and the additional metadata are fed back to the BS. At the BS side, the cluster components are individually decompressed as $\widehat{\widetilde{\mathbf{C}}}^{(\rm aln)}=f_{\rm de}(Q(f_{\rm en}(\widetilde{\mathbf{C}}^{(\rm aln)}))$. Based on \eqref{equ: align}, the original cluster $\mathbf{C}$ can also be derived from the aligned cluster $\widehat{\mathbf{C}}^{(\rm aln)}$ with
\begin{equation}
    \label{equ: align reverse}
    \mathbf{C}=\text{conj}(e^{-{\rm j}\beta_{t^\star}}\mathbf{S})\odot(\mathbf{F}_{\rm a}^{H}\widetilde{\mathbf{C}}^{\rm (aln)}\mathbf{F}_{\rm d}),
\end{equation}
where the peak positions are relocated and the peak phase is recovered. After the training of encoder and decoder NN modules, the decompressed cluster $\widehat{\widetilde{\mathbf{C}}}^{(\rm aln)}$ can approximate $\widetilde{\mathbf{C}}^{(\rm aln)}$. Note that the metadata $(n_1^\star,n_2^\star,m^\star,t^\star)$ is available at the BS, the original cluster $\widehat{\mathbf{C}}$ can also be reconstructed via \eqref{equ: align reverse}. Explicitly, the recovered cluster component $\widehat{\mathbf{C}}\in\mathbb{C}^{N_{\rm T}\times N_{\rm C}}$ in the spatial-frequency domain can be represented by 
\begin{equation}
    \label{equ: recovery}
    \widehat{\mathbf{C}}=\text{conj}(e^{-{\rm j}\beta_{t^\star}}\mathbf{S})\odot\left(\mathbf{F}_{\rm a}^{H}\widehat{\widetilde{\mathbf{C}}}^{(\rm aln)}\mathbf{F}_{\rm d}\right),
\end{equation}
where $\widetilde{\mathbf{C}}^{\rm (aln)}$ in \eqref{equ: align reverse} is substituted with $\widehat{\widetilde{\mathbf{C}}}^{\rm (aln)}.$ Denote the $l$th recovered cluster component as $\widehat{\mathbf{C}}_{l}$. Then, the reconstructed channel $\widehat{\mathbf{H}}$ in the spatial-frequency domain is yielded by summing up all recovered cluster components, i.e.,
\begin{equation}
    \label{equ: reconstruct channel}
    \widehat{\mathbf{H}}=\sum_{l=1}^{\widehat{R}}\widehat{\mathbf{C}}_{l}.
\end{equation}
The end-to-end inference runtime of EG-CsiNet is analyzed as follows. Once the multi-cluster decoupling is finished, the $\widehat{R}$ cluster components undergo parallel fine-grained alignment, encoder/decoder neural networks, and feature recovery modules in EG-CsiNet. Thus, stable end-to-end processing can be achieved. The details of practical runtime measurements are given in Sec.~\ref{subsec: sim results}.

\subsection{Overhead Analysis}
\label{subsec: feedback overhead}
As shown in the middle of Fig.~\ref{fig: model}, the feedback bits for $l$th decoupled cluster component comprises two parts, namely, the compression bits $\mathbf{b}^{(\rm f)}_{l}$ for aligned cluster component and the related metadata bits $\mathbf{b}^{(\rm m)}_{l}$. Owing to the oversampling-based scanning in \eqref{equ: peak position angular} and \eqref{equ: peak position delay}, the combinations of all possible peak positions are $O_{\rm a}N_{\rm T}O_{\rm d}N_{\rm C}$, where $O_{\rm a}=O_{\rm h}O_{\rm v}$ denotes the angular-domain oversampling factor. Consider both the size of the codebooks and the phase quantization bit $Q_{\rm p}$, the length of metadata bits is $q_{\rm m}=Q_{\rm p}+\lceil\log_{2}(O_{\rm a}N_{\rm T}O_{\rm d}N_{\rm C})\rceil$ for each decoupled cluster. Assume $Q_{\rm f}$-bit element-wise quantization is applied in the encoder module, the length of feedback bits for compressing each aligned cluster component is $q_{\rm f}=MQ_{\rm f}$. Based on the cluster-wise feedback manner in EG-CsiNet, the length of total feedback bits for a single channel instance is $\widehat{R}(q_{\rm m}+q_{\rm f})$. For the users distributed in a specific environment, the number of decoupled cluster components $\widehat{R}$ varies with the user location. In the practical uplink transmission of the feedback codeword, the number of decoupled clusters $\widehat{R}$ should be included at the beginning to facilitate adaptive feedback. Denoting the pre-defined maximal number of decoupled clusters as $R_{\max}$, then the feedback overhead for $\widehat{R}$ is $q_{R}=\lceil\log_{2}(R_{\max})\rceil$. Thus, the proposed EG-CsiNet can adaptively adjust the feedback overhead for each channel instance based on $\widehat{R}$, and the average feedback overhead is $q=q_{R}+\mathbb{E}\{\widehat{R}\}(q_{\rm m}+q_{\rm f})$ in a specific environment.

The workflow of the proposed EG-CsiNet is compatible with the standard CSI feedback workflow \cite{3gpp.38.214}. Akin to the standard CSI feedback, the workflow of EG-CsiNet involves six successive steps, including (1) configuration setup; (2) downlink pilot transmission; (3) CSI estimation; (4) compressed codewords and metadata calculation; (5) payload feedback; (6) decoding and CSI reconstruction. Intuitively, the proposed EG-CsiNet and the standard CSI feedback share the same number of BS-UE interactions within a single workflow. Since the proposed EG-CsiNet can robustly achieve better compression capability compared to the standard CSI feedback (see Sec.V-B for details), its feedback overhead can also be reduced under the same precision requirement. Thus, both the compatible design and strong compression capability of EG-CsiNet contribute to controlling the signaling and synchronization overhead. 

\subsection{Model Parameter Complexity}
\label{subsec: params complexity}
Benefiting from the cluster-wise feedback manner, different decoupled clusters can statistically reuse the model parameters, where the number of model parameters can be reduced under the same feedback overhead. Note that the $\widetilde{\mathbf{C}}^{(\rm aln)}$ has the same shape $N_{\rm T}\times N_{\rm C}$ with $\widetilde{\mathbf{H}}$. Additionally, both $\widetilde{\mathbf{C}}^{(\rm aln)}$ and $\widetilde{\mathbf{H}}$ exhibit sparse nature in the angular-delay domain. Thus, the NNs of the encoder and decoder module in the EG-CsiNet can adopt the same structure as the conventional deep learning-based CSI feedback algorithms, e.g., the CNN-based \cite{wcl_wen_2018_csinet, twc_guo_2020_csinetp} and transformer-based \cite{wcl_cui_2022_transnet} structures. Intuitively, the number of neural network parameters for the deep learning-based CSI feedback increases with the dimension $M$ of the compressed codeword. For the proposed EG-CsiNet, the dimension $M$ can be reduced since the encoder/decoder only compresses/reconstructs a single decoupled cluster. Without loss of generality, the encoder NN $f_{\rm en}$ is composed of the feature extraction module $f_{\rm en, ext}$ and a compressive linear module, i.e., $f_{\rm en}=f_{\rm en,lin}\circ f_{\rm en, ext}$ \cite{wcl_wen_2018_csinet,twc_guo_2020_csinetp,wcl_cui_2022_transnet}. Assume the compressed dimensions of the conventional AE-based CSI feedback and the proposed EG-CsiNet are $M_{1}$ and $M_{2}$, respectively. Then, under the same feedback overhead, i.e., $q_{\rm R}+\mathbb{E}\{\widehat{R}\}(M_{2}Q_{\rm f}+q_{\rm m})=M_1 Q_{\rm f}$, the compressed dimension ratio $M_1/M_2$ can be calculated by $M_{1}/M_{2}=\frac{q_{R}}{M_2 Q_{\rm f}}+\mathbb{E}\{\widehat{R}\}(1+\frac{q_{\rm m}}{M_2Q_{\rm f}})>\mathbb{E}\{\widehat{R}\}$. Thus, the parameter number of compressive linear module $f_{\rm en, lin}$ in the proposed EG-CsiNet can be reduced for more than $\mathbb{E}\{\widehat{R}\}$ times under the same feedback overhead. Similarly, the parameter of the decompression linear module in the decoder NN of EG-CsiNet can also be proportionally reduced by more than $\mathbb{E}\{\widehat{R}\}$ times. For the CNN-based model or the model with a higher encoding dimension $M$, the parameters of the compression and decompression linear modules dominate the total parameters of the model \cite{twc_guo_2020_csinetp}. Thus, the model parameters can be reduced in EG-CsiNet, which facilitates the deployment in memory-limited devices.

\section{Experiment and Discussion}
\subsection{Simulation Setup}
\label{subsec: setup}
In the simulations, two types of datasets are adopted to generate CSI data, which are detailed below.

\begin{figure}[t]
    \centering
    \vspace{-15pt}
    \subfloat{\includegraphics[width=0.24\textwidth]{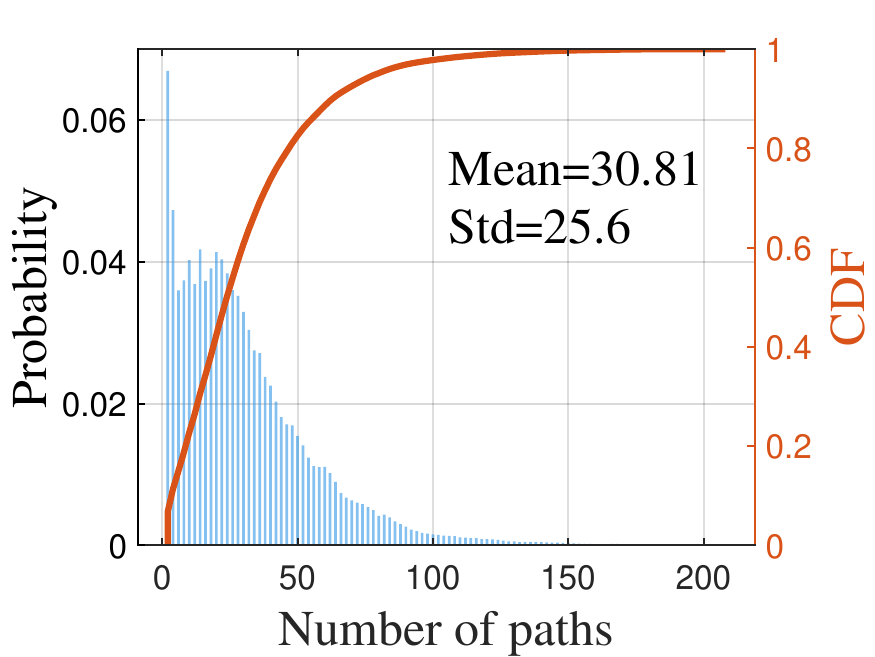}\label{subfig: path number}}
    \subfloat{\includegraphics[width=0.24\textwidth]{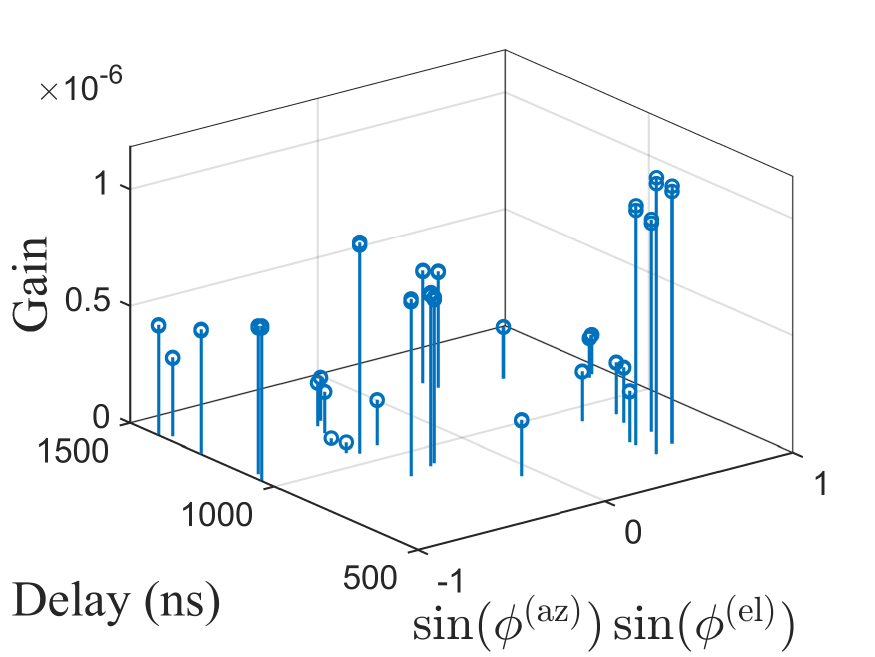}\label{subfig: params example}}
    \captionsetup{font=footnotesize}
    \caption{Illustration of the WAIR-D dataset. The distribution of the number of paths is shown on the left. Example of path parameters is shown on the right, where $\phi^{(\rm az)}/\phi^{(\rm el)}$ denotes the azimuth/elevation AoD.}
    \label{fig: WAIR-D illustration}
    \vspace{-15pt}
\end{figure}

\begin{enumerate}
    \item WAIR-D dataset \cite{huangfu2022wair}: This dataset is generated from the 3D ray-tracing tool built on 100 realistic maps from 40 major cities worldwide, which involves diverse building layouts. The carrier frequency is set as 2.6 GHz, and the path number is not clipped during the generation of CSI data. As shown on the left of Fig.~\ref{fig: WAIR-D illustration}, the average number of paths is 30.81. The path parameters in the dataset exhibit a clustered feature, where an example is shown on the right of Fig.~\ref{fig: WAIR-D illustration}.
    \item UMa dataset \cite{3gpp.38.901}: This dataset is generated under the specifications of 3GPP 38.901 document in UMa scenario. The height of the BS is set as 25 m, and the users are randomly distributed within a 120$^\circ$ sector with a 250 m radius. The carrier frequency is set as 3.5 GHz, and the indoor probability is set as 0.8. The number of clusters in LOS/NLOS scenarios is set as 12/20, where 20 paths are generated within a cluster. 
\end{enumerate}
A 32-antenna BS with a UPA is adopted in the datasets, where the numbers of horizontal and vertical antennas are set as 8 and 4, respectively. The number of UE antennas is set to 1. System bandwidth is set as 10 MHz, where the number of subcarriers is set as $N_{\rm C}=32$. In the simulation, up to 10 environments in the WAIR-D dataset and the UMa dataset are adopted for model pretraining, where the numbers of training samples are set as $9\times10^3$ per environment and $10^5$, respectively. To justify generalizability, the target dataset contains the other 90 environments in WAIR-D with a total of $9\times10^4$ samples. Considering the large dynamic range of pathloss, channel samples in the datasets are normalized by $\sqrt{N_{\rm T}N_{\rm C}}\mathbf{H}/\Vert\mathbf{H}\Vert_{F}$. By default, noise-free channel samples are assumed. 

The configurations for the proposed EG-CsiNet and the baselines are detailed as follows. Vanilla AE and UniversalNet+ \cite{wcnc_liu_2024_generalization} are adopted as the deep learning-based baselines. The vanilla AE refers to standard end-to-end training of the encoder and decoder NN without specialized generalization enhancement components, where the pretraining datasets include single-environment and mixed multi-environment \cite{twc_jiang_2024_multidomain}. Additionally, the enhanced type-II (eType-II) codebook \cite{3gpp.38.214} is adopted as a baseline without deep learning. Note that the proposed EG-CsiNet and deep learning-based baselines are not limited to specific neural network structures. To ensure a fair comparison, the proposed EG-CsiNet and the deep learning-based baselines employ the same neural network structure, with the standard CsiNet \cite{wcl_wen_2018_csinet} structure adopted by default. The quantization bit for the codeword is set as $Q_{\rm f}=6$ bits. The maximal number of decoupled clusters in EG-CsiNet is set as $R_{\max}=8$. The proposed EG-CsiNet and the deep learning-based baselines can adjust the dimension $M$ under different feedback overhead budgets. During the training, the Adam optimizer with an initial learning rate $10^{-3}$ is adopted, and the batch size is set as 64. The number of total training epochs is set to 200. The threshold in multi-cluster decoupling is set as $\eta=0.99$, and the oversampling factors are set as 2. The peak phase quantization bit is set as $Q_{\rm p}=2$ bits. The $\text{NMSE}=\mathbb{E}\{\Vert\widehat{\mathbf{H}}-\mathbf{H}\Vert_{F}^{2}/{\Vert\mathbf{H}\Vert_{F}^{2}}\}$ is adopted as the performance metric, and the proposed EG-CsiNet is targeted to achieve an NMSE at -18 dB in the target WAIR-D dataset. The performances of deep learning models are averaged over 3 random initializations. 

\subsection{Simulation Results}
\label{subsec: sim results}
\begin{figure}[t]
    \centering
    \captionsetup{font=footnotesize,justification=centering}
    \subfloat[Histogram of $\widehat{R}$]{\includegraphics[width=0.24\textwidth]{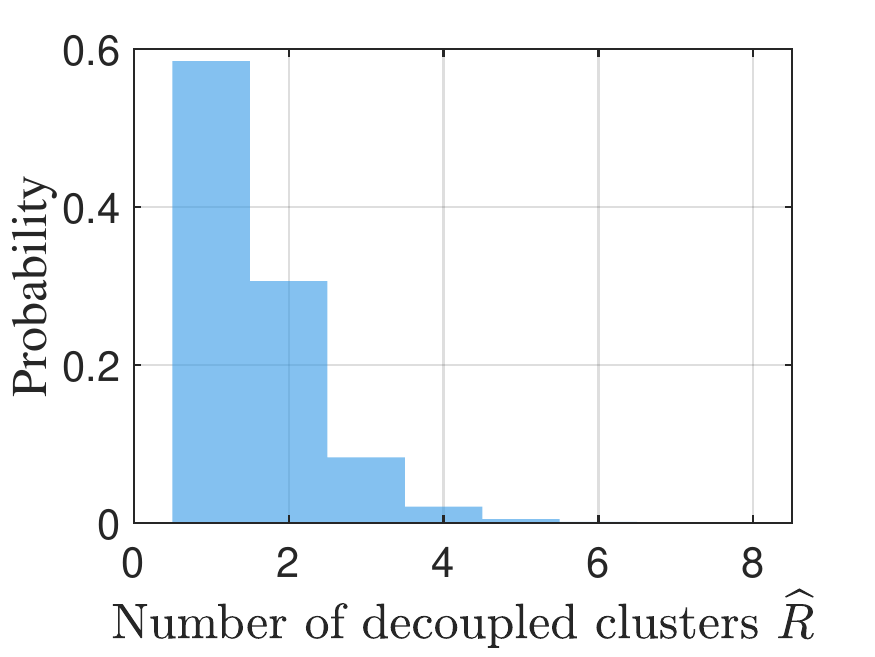}\label{subfig: cluster number}}\captionsetup{font=footnotesize,justification=centering}
    \subfloat[CDF of NMDE and UB-NPAE]{\includegraphics[width=0.24\textwidth]{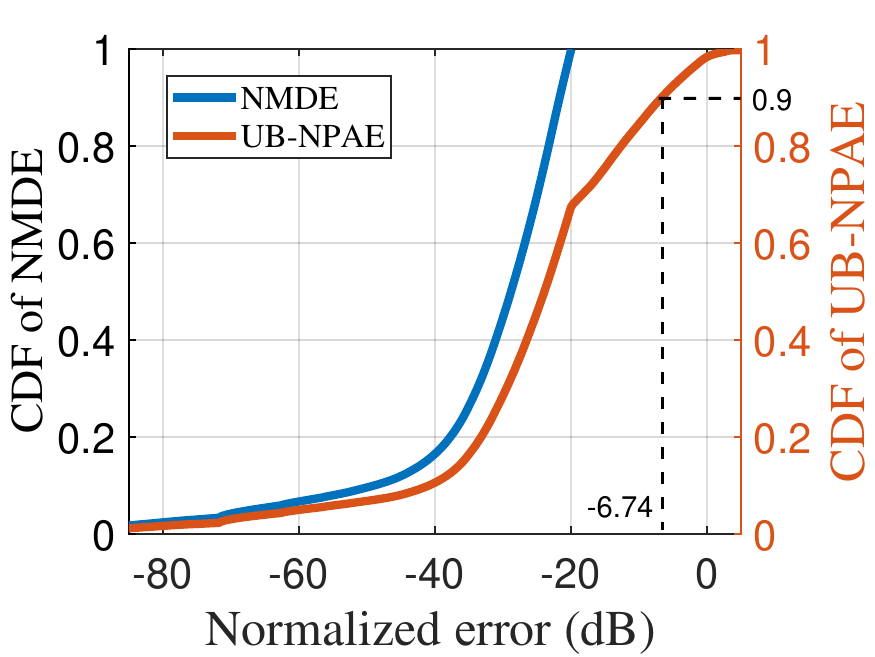}\label{subfig: NMDE UB-NPAE}}
    \captionsetup{font=small, justification=raggedright}
    \caption{Assessment of the decoupled clusters in the WAIR-D dataset.}
    \label{fig: cluster decoupling}
    \vspace{-15pt}
\end{figure}

\subsubsection{Assessment of Multi-Cluster Decoupling}

To assess the accuracy of the SVD-based multi-cluster decoupling, the NMDE and the upper-bound of normalized physical-association error (UB-NPAE) are adopted \cite{wang2025generalizable}. Intuitively, a low NMDE indicates that the impact of the undetected physical paths on the channel feedback precision is negligible. UB-NPAE is adopted to quantify the accuracy of decoupled clusters $\{\mathbf{C}^{\star}_{l}\}_{l=1}^{\widehat{R}}$ comparing to the ground-truth physical paths and is defined as 
\begin{equation}
    \label{equ: UB-NPAE}
    \text{UB-NPAE}\!=\!\frac{\sum_{l=1}^{\widehat{R}}\!\Vert\mathbf{C}^{\star}_{l}\!-\!\!\sum_{(l^{\prime}\!,i)\in\mathcal{K}_{l}^{\star}}\mathbf{A}_{l^{\prime}\!,i}\Vert_{F}^2}{\Vert\mathbf{H}\Vert_{F}^2},
\end{equation}
where $\mathbf{A}_{l^{\prime},i}\in\mathbb{C}^{N_{\rm T}\times N_{\rm C}}$ denotes the response $i$th physical path from the $l^\prime$th cluster, $\{\mathcal{K}_{l}^{\star}\}_{l=1}^{\widehat{R}}$ denotes a weak partition of the index set $\mathcal{S}=\{(l,i)|1\leq l\leq N_{\rm cl},1\leq i\leq N_{{\rm p},l}\}$ such that $\mathcal{K}_{l_{1}}^\star\cap\mathcal{K}_{l_{2}}^\star=\emptyset,~\bigcup_{l=1}^{\widehat{R}}\mathcal{K}_{l}^\star=\mathcal{S}$ and $\mathcal{K}_{l}^\star$ can be empty, where the derivations can be found in the Appendix~A of \cite{wang2025generalizable}. Firstly, the number of the decoupled clusters $\widehat{R}$ in the WAIR-D dataset is illustrated on the left of Fig.~\ref{fig: cluster decoupling}. It can be found that $\widehat{R}$ is far less than the number of the physical paths on the left of Fig.~\ref{fig: WAIR-D illustration}, which verifies the low-rank property for the cluster-based channel. Secondly, the cumulative density functions (CDFs) of NMDE and UB-NPAE are illustrated on the right of Fig.~\ref{fig: cluster decoupling}. It can be found that the NMDE of all the samples is less -20 dB by setting the threshold $\eta$ in \eqref{equ: threshold} as 0.99. Additionally, the average NMDE is -26.2 dB, which is far less than the target NMSE at -18 dB. Hence, the rationality to choose $\eta$ is validated. Moreover, the 90th percentile of UB-NPAE is -6.74 dB (See the dashed line in the right of Fig.~\ref{fig: cluster decoupling}.), indicating that the decoupled clusters $\{\mathbf{C}_{l}^\star\}_{l=1}^{\widehat{R}}$ can largely reflect the structure of physical paths in the channel $\mathbf{H}$. 

\begin{figure}[t]
        \centering
        \includegraphics[width=0.5\textwidth]{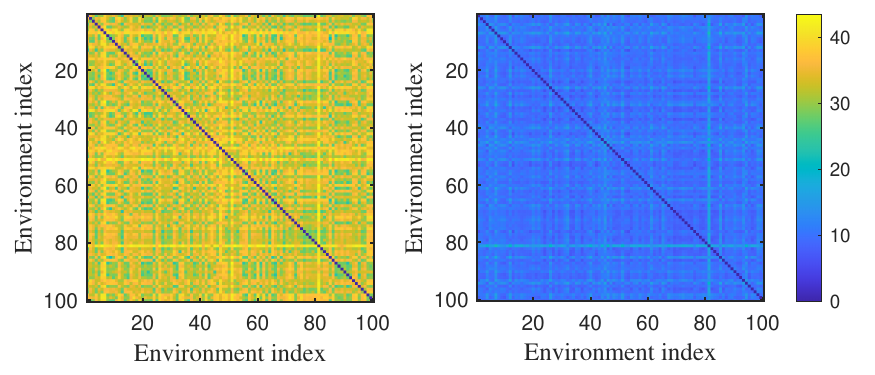}
    \captionsetup{font=footnotesize}
    \caption{Wasserstein-1 distance heatmap of the original channel (left) and the aligned cluster (right) in the WAIR-D dataset. }
    \label{fig: wasserstein}
    \vspace{-15pt}
\end{figure}

\subsubsection{Generalizability Comparison}
\begin{figure}[t]
        \centering
        \includegraphics[width=0.45\textwidth]{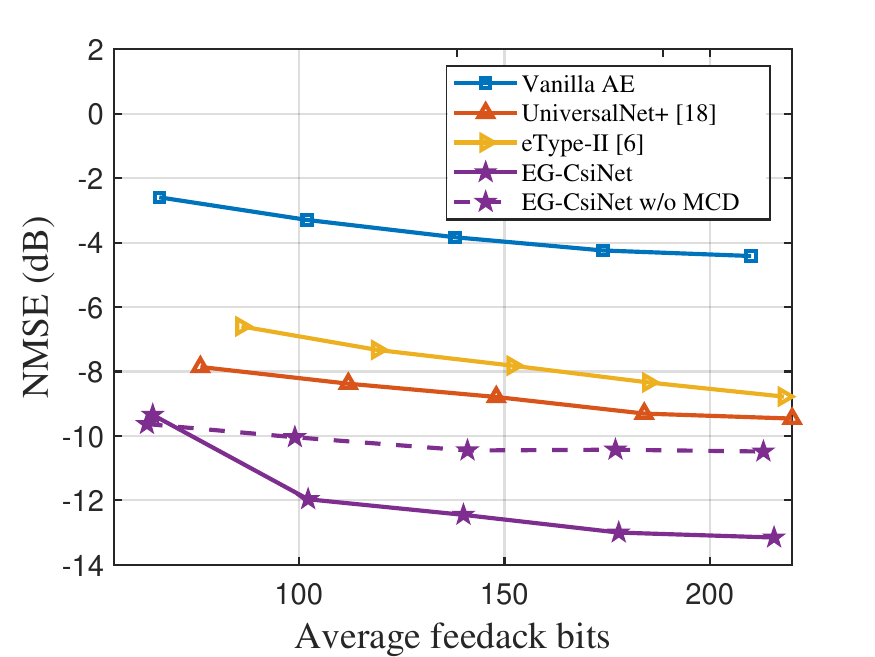}
    \captionsetup{font=footnotesize}
    \caption{Generalization NMSE under different feedback overhead, where the encoder and decoder are pretrained in a single environment. }
    \label{fig: generalizability-1}
    \vspace{-15pt}
\end{figure}
Firstly, the performance of the physics-based distribution alignment is investigated. Here, the Wasserstein-1 distance \cite[Definition 6.1]{villani2009optimal} is adopted as a metric to quantify the distribution alignment performance in the 100 environments of WAIR-D, which is plotted in Fig.~\ref{fig: wasserstein}. For the original channel in WAIR-D, the average cross-environment Wasserstein-1 distance is 33.82, which indicates obvious distribution shifts. By applying the physics-based distribution alignment, the average cross-environment Wasserstein-1 distance of the aligned cluster is reduced to 10.39. Leveraging the Kantorovich-Rubinstein duality \cite{villani2009optimal}, the cross-environment reconstruction error of the trained autoencoder is bounded and scales with the Wasserstein-1 distance between the environmental distributions. Since neural networks can effectively capture angular-delay domain correlations, the autoencoder can achieve a low training reconstruction error. Therefore, a low reconstruction error in the test environment can also be achieved, which can fundamentally enhance the model generalizability. 

\begin{figure}[t]
    \vspace{-10pt}
        \centering
        \includegraphics[width=0.45\textwidth]{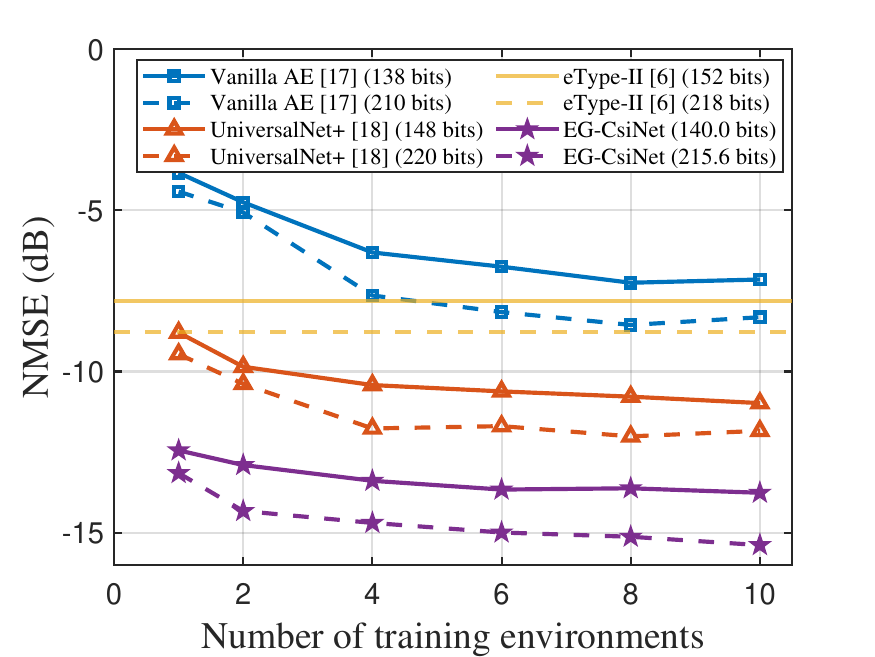}
    \captionsetup{font=footnotesize}
    \caption{Generalization NMSE with varying number of training environments in WAIR-D, where the average feedback bits are shown in the legend.}
    \label{fig: generalizability-3}
    \vspace{-10pt}
\end{figure}
Next, a single pretraining environment in WAIR-D is utilized, and the generalization NMSE in the 90 unseen environments is plotted in Fig.~\ref{fig: generalizability-1}. It can be found that the NMSE of the proposed EG-CsiNet has been reduced by more than 3.5 dB compared to the baselines in unseen environments. Compared to UniversalNet+, the proposed EG-CsiNet can better address the distribution shift of CSI samples across different environments. Moreover, due to the drastic CSI distribution shift, the vanilla AE cannot achieve accurate channel feedback in unseen environments with a single source environment, where the generalization NMSE has been degraded for 8 dB compared to the proposed EG-CsiNet. To further investigate the mechanism for generalizability enhancement, we also consider the EG-CsiNet without the multi-cluster decoupling module, which is plotted as the dashed line in Fig.~\ref{fig: generalizability-1}. Comparing the EG-CsiNet with its counterpart without multi-cluster decoupling (MCD), it can be found that the generalization NMSE of EG-CsiNet can be reduced by more than 2.5 dB with the increase of average feedback bits. The rationale lies in the fact that multi-cluster decoupling can effectively address the distribution shift of the multi-cluster structure. Moreover, comparing the EG-CsiNet without multi-cluster decoupling and vanilla AE, it can be found that the generalization NMSE can be reduced by 6 dB with the fine-grained alignment module. Thus, the multi-cluster decoupling and fine-grained alignment modules in EG-CsiNet can effectively address the cross-environment distribution shift, which validates Sec.~\ref{subsec: address shift}.

Then, the environment-generalizability with a varying number of training environments in WAIR-D is investigated, which is depicted in Fig.~\ref{fig: generalizability-3}. It can be found that the proposed EG-CsiNet still achieves the best channel feedback performance in unseen environments with multiple training environments compared to the baselines, where the generalization NMSE can be further reduced by more than 3 dB. Additionally, the environment-generalizability of the proposed EG-CsiNet can also be gradually improved with an increasing number of training environments, which also facilitates its deployment under different availability of training data sources.  

\begin{figure}[t]
        \centering
        \includegraphics[width=0.45\textwidth]{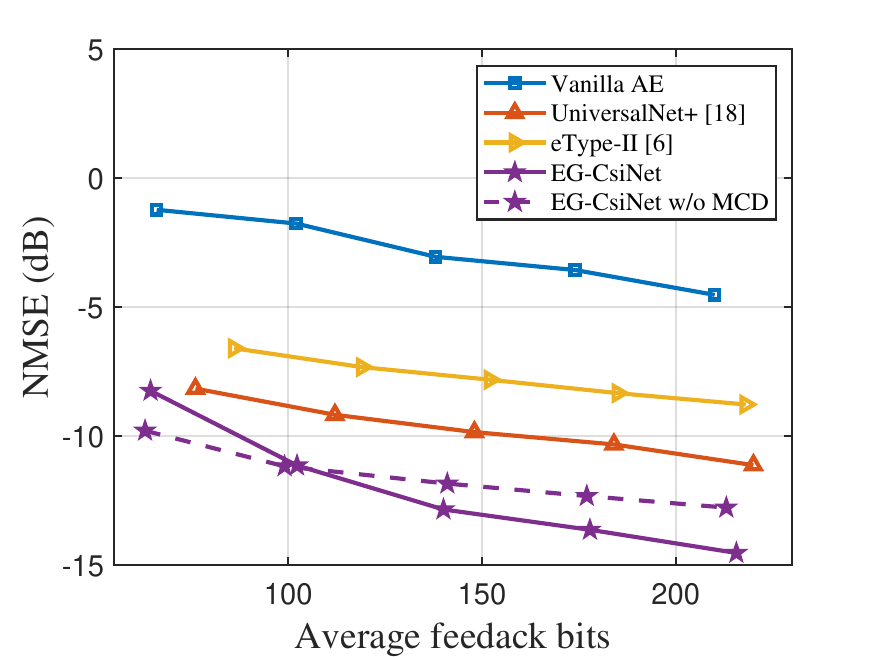}
    \captionsetup{font=footnotesize}
    \caption{Generalization NMSE under different feedback overhead, where the deep learning-based models are pretrained in the UMa dataset.}
    \label{fig: pretrain in UMa}
    \vspace{-10pt}
\end{figure}

Further, the environment generalization with different pretraining dataset types is examined. Explicitly, the UMa dataset is adopted for pretraining, and the generalization comparison of the proposed EG-CsiNet and baselines in the 90 unseen environments of WAIR-D is plotted in Fig.~\ref{fig: pretrain in UMa}. Compared to the baselines, the proposed EG-CsiNet can still achieve the best generalization performance in the unseen environments, where the generalization NMSE has been reduced for 3 dB in under the same feedback overhead level. Thus, the distribution of WAIR-D and UMa datasets can be effectively aligned in the proposed EG-CsiNet. When the feedback overhead budget is sufficient ($>$100 bits), the proposed EG-CsiNet in Fig.~\ref{fig: pretrain in UMa} can still achieve the best CSI reconstruction precision compared to its ablation without multi-cluster coupling, which is consistent with the result in Fig.~\ref{fig: generalizability-1}.

\begin{figure}[t]
    \vspace{-5pt}
    \centering
    \subfloat{\includegraphics[width=0.24\textwidth]{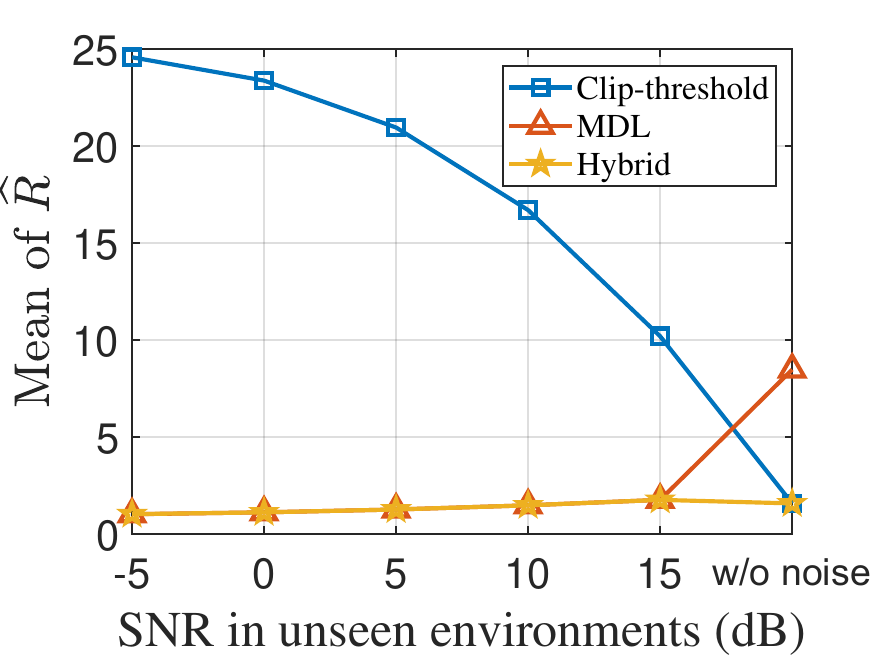}\label{subfig: mean of R}}
    \subfloat{\includegraphics[width=0.24\textwidth]{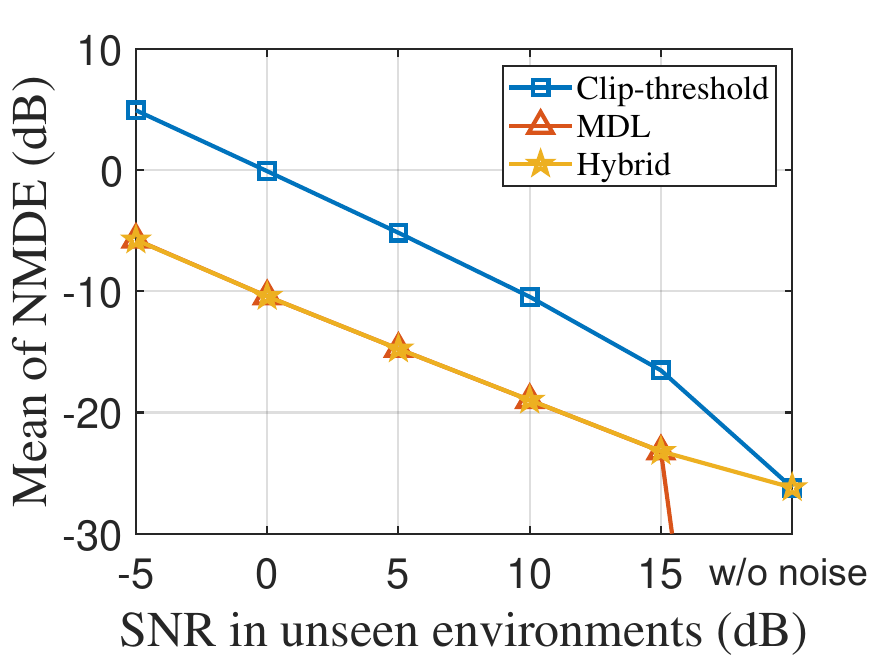}\label{subfig: mean of NMDE}}
    \captionsetup{font=footnotesize}
    \caption{Robustness of proposed multi-cluster decoupling against different levels of channel estimation error. 
    }
    \label{fig: decoupling robustness}
    \vspace{-15pt}
\end{figure}
\begin{figure*}[t]
    \centering
    \captionsetup{font=footnotesize,justification=centering}
    \subfloat[Pretrained in 1 environment of WAIR-D]
    {\includegraphics[width=0.32\textwidth]{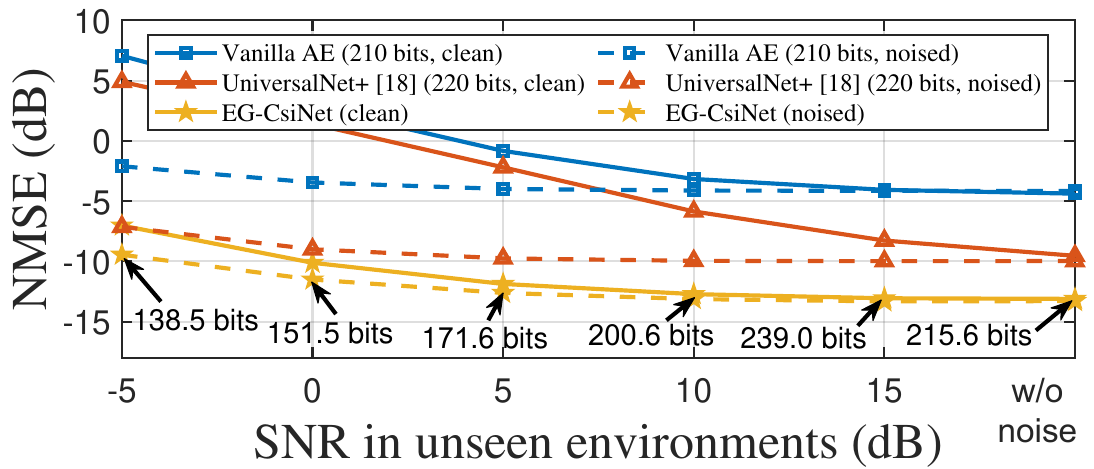}\label{subfig: pretrain 1 env noise}}
    \captionsetup{font=footnotesize,justification=centering}
    \subfloat[Pretrained in 10 environments of WAIR-D]
    {\includegraphics[width=0.32\textwidth]{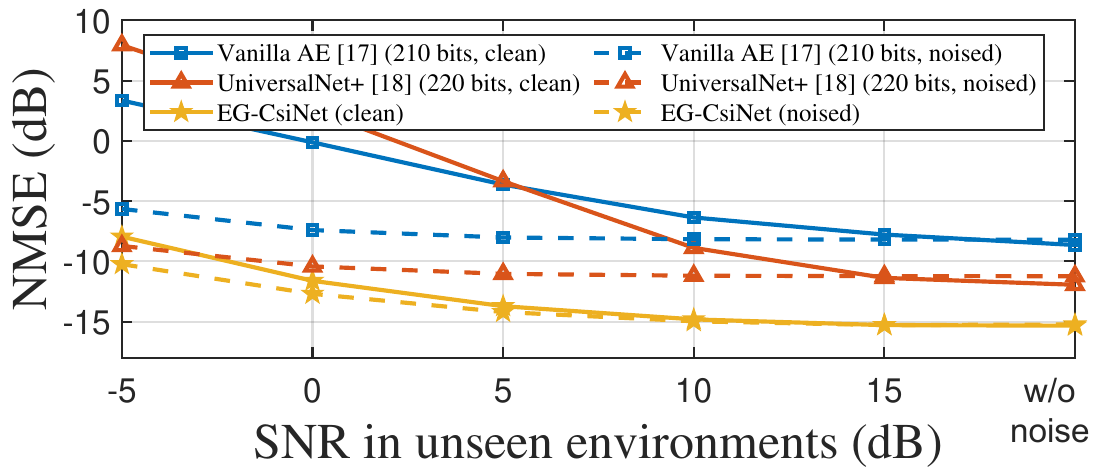}\label{subfig: pretrain 10 envs noise}}
    \captionsetup{font=footnotesize,justification=centering}
    \subfloat[Pretrained in UMa dataset]
    {\includegraphics[width=0.32\textwidth]{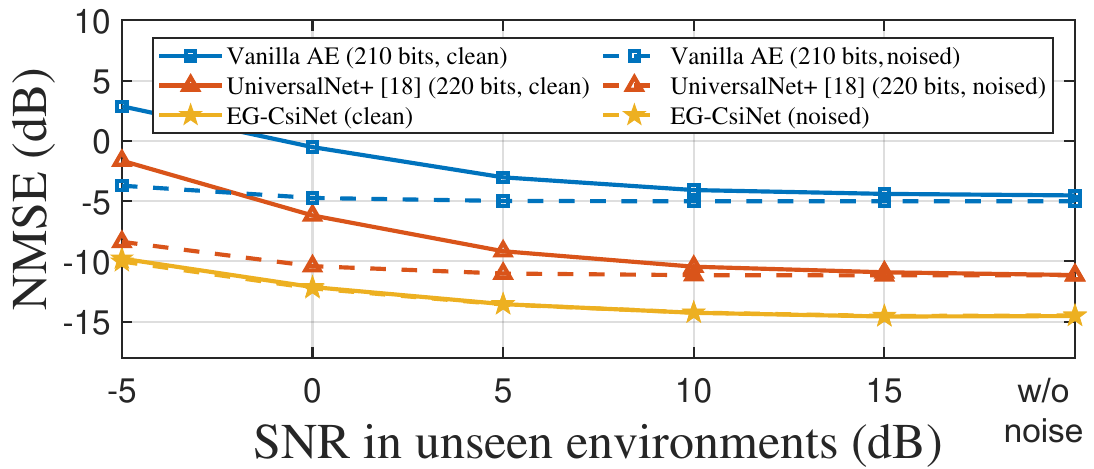}\label{subfig: pretrained in UMa noise}}\captionsetup{font=small,justification=raggedright}
    \caption{Generalization comparison with channel estimation error. (Average feedback bits of EG-CsiNet are marked in the first plot.)}
    \label{fig: noise robustness}
    \vspace{-5pt}
\end{figure*}

\begin{figure*}[t]
    \centering
    \captionsetup{font=footnotesize,justification=centering}
    \subfloat[Pretrained in 1 environment of WAIR-D]
    {\includegraphics[width=0.32\textwidth]{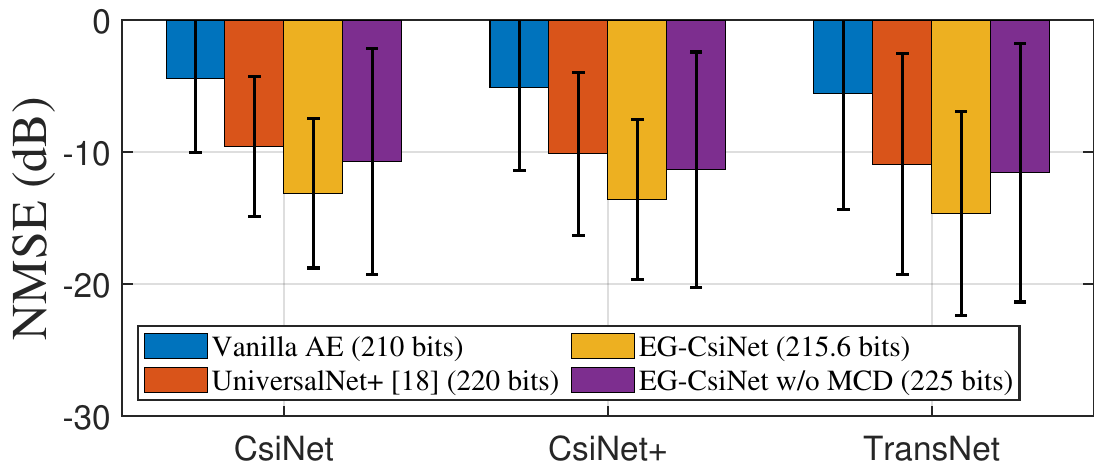}\label{subfig: pretrain 1 env}}
    \captionsetup{font=footnotesize,justification=centering}
    \subfloat[Pretrained in 10 environments of WAIR-D]
    {\includegraphics[width=0.32\textwidth]{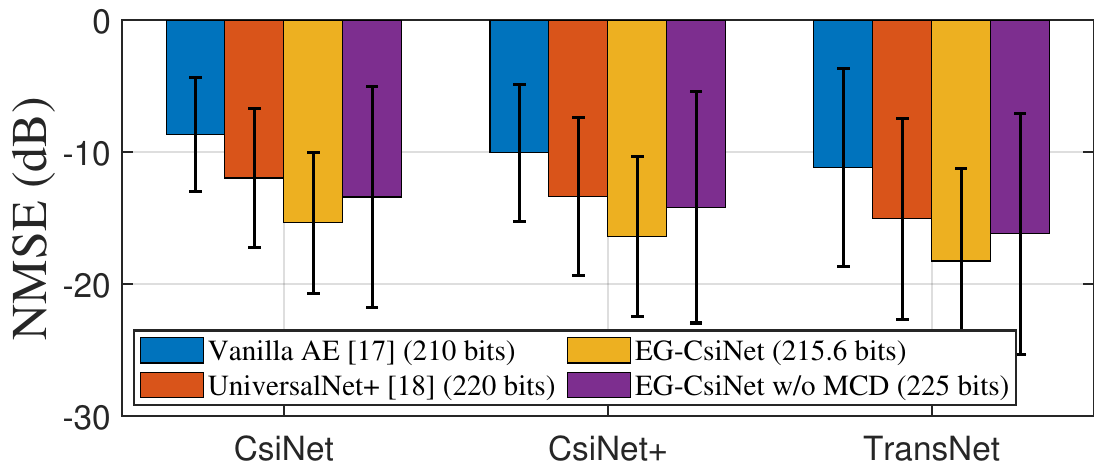}\label{subfig: pretrain 10 envs}}
    \captionsetup{font=footnotesize,justification=centering}
    \subfloat[Pretrained in UMa dataset]
    {\includegraphics[width=0.32\textwidth]{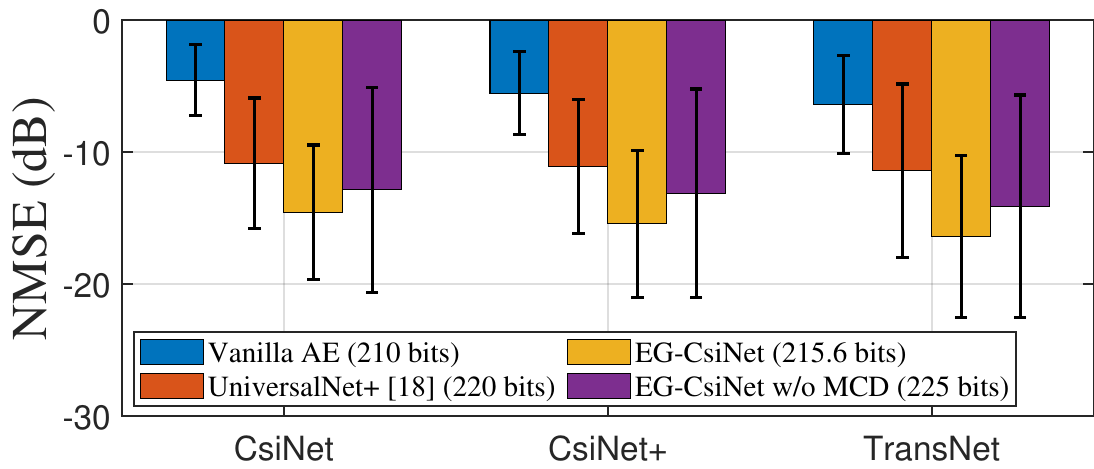}\label{subfig: pretrained in UMa}}\captionsetup{font=small,justification=raggedright}
    \caption{Generalization comparison with different NN structures, where the error bars denote the standard deviations of the NMSE in dB. }
    \label{fig: NN structure}
    \vspace{-10pt}
\end{figure*}
\subsubsection{Robustness Against Channel Estimation Error}

The noise-robustness of the proposed EG-CsiNet is justified as follows. Here, SNR in estimated channel $\mathbf{H}^{(\rm e)}$ is defined as $\rm{SNR}=\Vert\mathbf{N}^{(\rm e)}\Vert_{F}^2/\Vert\mathbf{H}\Vert_{F}^2$. Firstly, the denoising capability of the multi-cluster decoupling is verified. Under different criteria for cluster number estimation, the average number of decoupled clusters $\widehat{R}$ is plotted on the left of Fig.~\ref{fig: decoupling robustness}, while the mean of NMDE is plotted on the right. Intuitively, an optimal tradeoff between the number of decoupled clusters and the NMDE can be achieved under the hybrid criterion, which is essential for robust CSI feedback. On the contrary, when only the MDL criterion is adopted, the number of decoupled clusters will be overestimated for the noise-free condition, which increases the feedback overhead. Meanwhile, the noise components in the low-SNR regime cannot be filtered when only the clip-threshold criterion is adopted. Then, the pretrained models are tested in the unseen environments with different levels of ${\rm SNR}$. Here, we consider two pretraining strategies concerning the channel estimation error: (1) the models are pretrained on a clean CSI dataset without estimation error; (2) the models are pretrained on a noised CSI dataset, in which estimation errors ranging from -5 to 15 dB are introduced into the input, while the corresponding noise-free CSI serves as the supervision labels. Then, the generalization NMSE under different $\rm{SNR}$ is plotted in Fig.~\ref{fig: noise robustness}. With clean pretraining datasets, the proposed EG-CsiNet exhibits obvious generalization gain compared to the baselines, especially in the low-SNR regime. The rationale lies in the fact that the SVD-based multi-cluster decoupling exhibits a strong denoising capability in the presence of channel estimation error, which guarantees the robustness of the EG-CsiNet. On the contrary, the generalization performance of baselines degrades in the low-SNR regime due to the distribution shifts caused by channel estimation error. With the noised pretraining datasets, the generalization NMSE of the proposed EG-CsiNet can still be lowered by $2\sim3$ dB compared to the UniversalNet+, which further justifies the robust generalization gain. 

\subsubsection{Flexibility on NN Structures}
Further, environment-generalizability comparison with different NN structures is investigated, which includes the CNN-based CsiNet/CsiNet+ and the transformer-based TransNet. Here, the compression dimension $M$ of the EG-CsiNet and the baselines (including UniversalNet+ and vanilla AE) is set as 20 and 35 to guarantee the same feedback overhead level (210$\sim$220 bits). Then, the generalizability comparisons under different NN structures and pretraining datasets are illustrated in Fig.~\ref{fig: NN structure}. \footnote{Note that different NN structures exhibit varying compression capability, which also impacts the reconstruction NMSE of the proposed EG-CsiNet, vanilla-AE, and UniversalNet+ in the unseen environments.} It can be found that the generalization NMSE of the proposed EG-CsiNet can be robustly reduced by $3\sim5$ dB compared to the SOTA, which is compatible with various NN structures. Additionally, it can be observed that the EG-CsiNet with a small-sized CsiNet structure can achieve a lower NMSE compared to the baselines with a large-sized TransNet structure, which also facilitates deployment with limited memory resources. Comparing EG-CsiNet with its ablation without MCD, it can be found that the generalization NMSE of the proposed EG-CsiNet can be reduced by $1.7\sim3.1$dB. Meanwhile, the standard deviation of generalization NMSE in EG-CsiNet can also be reduced by $2.1\sim3.1$dB with MCD. Thus, the proposed EG-CsiNet can benefit from MCD to achieve generalizability and robustness across different training datasets and specific neural network structures. 

\begin{table}[t]
  \centering
  \belowrulesep=0.5pt
  \aboverulesep=0pt
  \captionsetup{font=footnotesize}
  \caption{Neural network parameter comparison under the same feedback overhead level (210$\sim$220 bits)}
    \begin{tabular}{c|c|c|c|c}
    \toprule
    \multirow{2}{*}{NN~structure} & \multicolumn{2}{c|}{Encoder module} & \multicolumn{2}{c}{Decoder module} \\
\cmidrule{2-5}          & EG-CsiNet & Baselines & EG-CsiNet & Baselines \\
    \midrule
    CsiNet \cite{wcl_wen_2018_csinet} & 41,022 & 71,757 & 46,332 & 77,052 \\
    \midrule
    CsiNet+ \cite{twc_guo_2020_csinetp} & 41,380 & 72,115 & 64,828 & 95,548 \\
    \midrule
    TransNet \cite{wcl_cui_2022_transnet} & 322,132 & 352,867 & 340,928 & 371,648 \\
    \bottomrule
    \end{tabular}%
  \label{tab:parameter}%
  \vspace{-10pt}
\end{table}%

\subsubsection{Complexity Comparison} The neural network parameter reduction in the proposed EG-CsiNet is investigated. Here, the neural network parameter comparison under the same feedback overhead is presented in Table~\ref{tab:parameter}. For the CsiNet and CsiNet+ structures, the number of encoder parameters in EG-CsiNet is reduced by 42.8\% and 42.6\%, respectively, while the number of decoder parameters is reduced by 39.8\% and 32.15\%, respectively. The rationale lies in the dominance of the compression and decompression linear modules in the total parameters of the CNN-based CsiNet/CsiNet+ \cite{twc_guo_2020_csinetp}. For the TransNet structure, the number of encoder and decoder parameters in EG-CsiNet is reduced by 8.7\% and 8.3\%, where the parameter number of the compression and decompression linear modules is less dominant in the total parameters compared to the CsiNet/CsiNet+. 

\begin{figure}[t]
    \vspace{-5pt}
    \centering
    \captionsetup{font=footnotesize,justification=centering}
    \subfloat[Runtime comparison between EG-CsiNet and the baselines]{\includegraphics[width=0.24\textwidth]{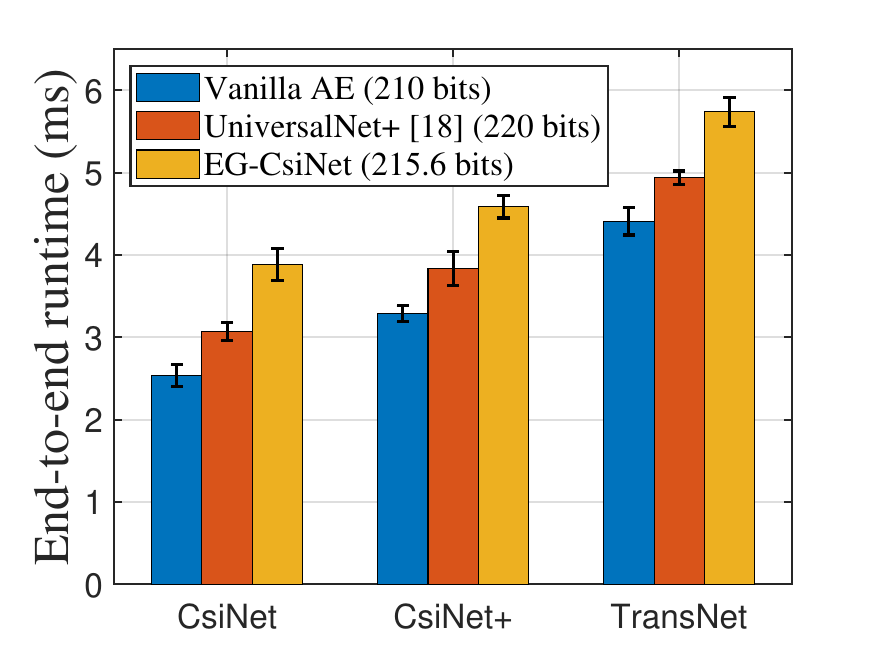}\label{subfig: e2e}}
    \captionsetup{font=footnotesize,justification=centering}
    \subfloat[Runtime of EG-CsiNet under different $\widehat{R}$]{\includegraphics[width=0.24\textwidth]{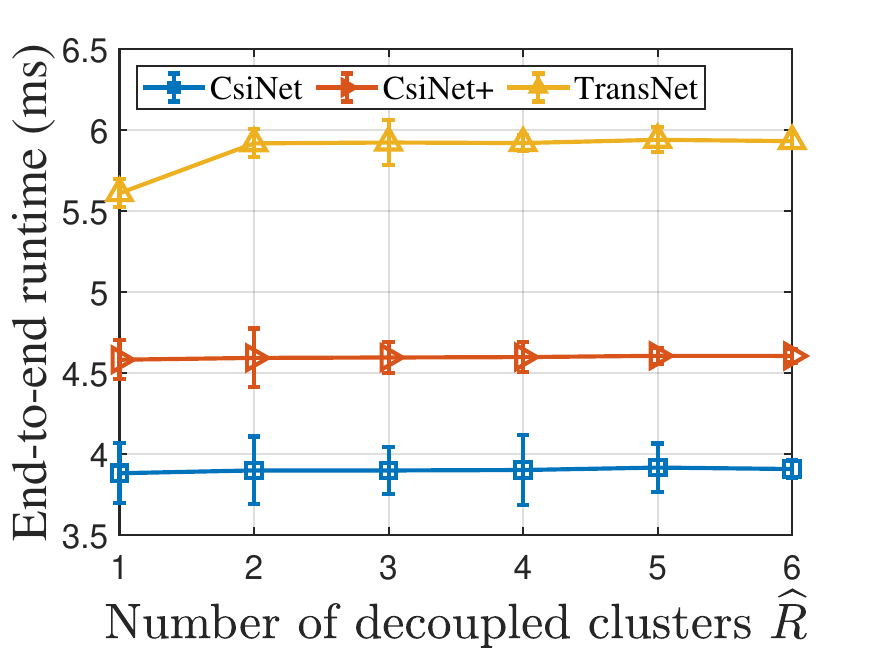}\label{subfig: versus paths}}
    \captionsetup{font=small,justification=raggedright}
    \caption{Comparison of practical end-to-end runtime.}
    \label{fig: runtime}
    \vspace{-15pt}
\end{figure}

Next, the runtime of the proposed EG-CsiNet is thoroughly investigated, which is vital for real-time CSI feedback in practical scenarios. Here, the inference runtime is measured on the Nvidia GeForce RTX 3090 device, where the batch size is set as 1. Firstly, the end-to-end runtime of vanilla AE, UniversalNet+, and the proposed EG-CsiNet under different NN structures is investigated, which is plotted on the left of Fig.~\ref{fig: runtime}. It can be found that the practical end-to-end runtime of the proposed EG-CsiNet can be controlled within certain milliseconds for different NN structures, which is applicable for real-time CSI feedback. The runtime increase of the proposed EG-CsiNet is around 1.3 ms compared to the vanilla AE baseline, which is relatively low. Specifically, the average runtime of the SVD-based multi-cluster decoupling in EG-CsiNet is only 0.4 ms, which is efficiently executed. The rationale lies in the fact that the proposed SVD-based multi-cluster decoupling directly leverages the cluster-level property, which avoids the time-consuming intermediate path parameter estimation. Further, the end-to-end runtime of EG-CsiNet under different numbers of decoupled clusters $\widehat{R}$ is investigated, which is plotted on the right of Fig.~\ref{fig: runtime}. It can be observed that the end-to-end runtime of EG-CsiNet remains under different numbers of decoupled clusters, which stems from the parallel computing mechanism of the proposed EG-CsiNet and facilitates stable processing. Additionally, the end-to-end runtime of the proposed EG-CsiNet under different numbers of antennas is investigated. When the number of BS antennas is set as 32, 64, and 128, the end-to-end runtime of the proposed EG-CsiNet is 3.8 ms, 4.0 ms, and 4.5 ms. Thus, the runtime increase of the proposed EG-CsiNet is less than 1 ms when the antenna number is increased from 32 to 128, which justifies the applicability of EG-CsiNet to larger antenna arrays. 

\subsection{Sim-to-Real Results}
\label{subsec: sim-to-real}

\begin{figure}[t]
    \vspace{-10pt}
    \centering
    \captionsetup{font=footnotesize,justification=centering}
    \subfloat[Sim-to-real generalization NMSE with different feedback bits]{\includegraphics[width=0.45\textwidth]{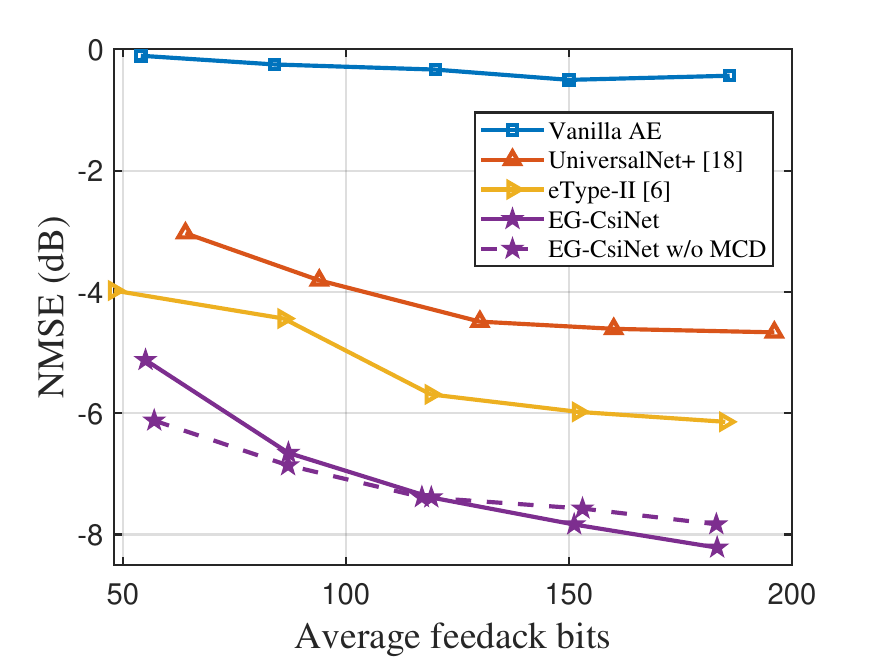}\label{subfig: sim2real bits}}
    \captionsetup{font=footnotesize,justification=centering}
    \subfloat[Sim-to-real generalization NMSE with different NN structures]{\includegraphics[width=0.45\textwidth]{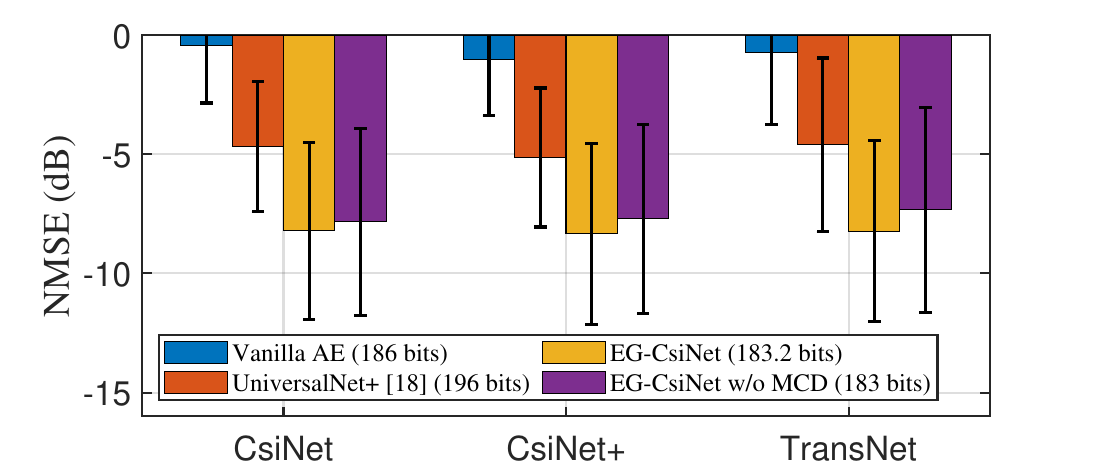}\label{subfig: sim2real NN}}
    \captionsetup{font=small,justification=justified}
    \caption{Sim-to-real generalization comparison over real-world RENEW dataset \cite{tvt_du_2022_dataset}.}
    \label{fig: sim2real}
    \vspace{-15pt}
\end{figure}

In this subsection, a sim-to-real experiment is conducted to justify the generalizability of the deep learning models in real-world massive MIMO systems. Explicitly, the models are pretrained in a simulated channel dataset and are directly tested with real-world channel measurements without any fine-tuning, which is a critical and challenging evaluation for real-world applications. Here, the UMa dataset in Sec.~\ref{subsec: setup} is utilized as the pretraining dataset, and the real-world massive MIMO channel measurements dataset from the RENEW project is adopted as the target dataset \cite{tvt_du_2022_dataset}, which is denoted as the RENEW dataset. The RENEW dataset was measured in a campus environment, including channel samples collected from 4 LOS areas and 5 NLOS areas. To align the system dimensional parameter settings with the pretraining dataset, the target dataset is yielded by extracting the $8\times 4$ sub-array and the first 32 subcarriers from the original RENEW dataset. Then, the sim-to-real generalization NMSE under different feedback overhead is shown Fig.~\ref{fig: sim2real}\subref{subfig: sim2real bits}. Under the same feedback overhead level, the sim-to-real generalization NMSE of EG-CsiNet can be reduced by 3.5 dB and 7.5 dB compared to the UniversalNet+ and vanilla AE. Thus, the proposed EG-CsiNet exhibits strong sim-to-real generalizability, which facilitates the large-scale deployment of the pretrained models. Further, the sim-to-real generalization comparison with different NN structures is plotted in Fig.~\ref{fig: sim2real}\subref{subfig: sim2real NN}, where the average feedback bits are set at the level of 180$\sim$196 bits for fair comparison. It can be found that the proposed EG-CsiNet can robustly achieve the best sim-to-real generalization performance under different NN structures, where the applicability is further justified in the real world. 

\subsection{Future Work Discussion}
\subsubsection{Intelligent feedback overhead allocation} As depicted in Fig.~\ref{fig: pretrain in UMa} and Fig.~\ref{fig: sim2real}\subref{subfig: sim2real bits}, a performance drop of EG-CsiNet compared to the ablation without multi-cluster decoupling can be found when the feedback overhead budget is less than 100 bits. The reason lies in the equal feedback overhead allocation to each decoupled cluster. Practically, the power distribution of the decoupled clusters is imbalanced, and their contribution to the overall channel reconstruction precision is not equal. Therefore, equal feedback overhead allocation will limit the compressed dimension with stronger power and degrade the overall reconstruction precision, which is more obvious with a small feedback overhead budget. To this end, an intelligent feedback overhead allocation scheme should be designed to minimize the overall reconstruction error under the feedback overhead budget. Meanwhile, the mutli-resolution encoder and decoder should be designed to facilitate the overhead allocation scheme as well.

\subsubsection{Multi-cluster decoupling with soft constraints}As presented in \eqref{equ: cluster optimization}, current multi-cluster decoupling optimization in EG-CsiNet is formulated with hard constraints $\mathcal{C}_1$ and $\mathcal{C}_2$. Although hard constraints $\mathcal{C}_1$ and $\mathcal{C}_2$ facilitate a closed-form solution via the EYM theorem, these approximations may not universally be satisfied. To this end, the multi-cluster decoupling can be optimized with modified soft constraints. For instance, the constraint $\mathcal{C}_1$ can be modified into a soft constraint 
\begin{equation}
    \label{equ: rank-one soft}
    \mathcal{C}_1^\prime: \sigma_{1,l}^2/\Vert\mathbf{C}_{l}\Vert_F^2\geq1-\epsilon_1,~~1\leq l\leq\widehat{R},
\end{equation}
where $\sigma_{1,l}$ denotes the largest singular value of $\mathbf{C}_{l}$ and $\epsilon_1$ denotes a predefined small constant. Additionally, the dual-orthogonality constraint $\mathcal{C}_2$ can be modified as a soft constraint:
\begin{equation}
    \label{equ: orthogonality soft}
    \mathcal{C}_2^\prime: \Vert\mathbf{C}_{l}\mathbf{C}_{l^\prime}^{H}\Vert_{F}\leq\epsilon_2,\quad\Vert\mathbf{C}_{l}^{H}\mathbf{C}_{l^\prime}\Vert_{F}\leq\epsilon_2,\quad\quad \forall l\neq l^{\prime},
\end{equation}
where $\epsilon_2$ is a predefined small constant to control the angular and delay domain inter-cluster correlation. By modifying the hard constraints $\mathcal{C}_1$ and $\mathcal{C}_2$ into soft constraints, the optimization of multi-cluster decoupling can be better physically grounded. Different from  \eqref{equ: cluster optimization}, multi-cluster decoupling with soft constraints does not have a closed-form solution, where an efficient algorithm deserves further investigation. 

\subsubsection{Real-time feasibility on commercial device}In Fig.~\ref{fig: runtime}, the end-to-end runtime results are measured on the device Nvidia GeForce RTX 3090, which can give a fair comparison between the EG-CsiNet and the baselines. In commercial systems, the hardware of the user and the BS are separated, and a more practical end-to-end runtime evaluation is required. Currently, the leading vendors in the industry are actively developing enhanced computational support for DL models \cite{commag_Kundu_2025_AI-RAN}. Thus, an end-to-end real-time feasibility evaluation on the DL-enabled commercial mobile platforms is left as our future work. 

\subsubsection{Practical channel modeling with non-idealities} Comparing the results in Fig.~\ref{fig: sim2real} and the Fig.~\ref{fig: NN structure}, it can be observed that the generalization gain of MCD in the challenging sim-to-real experiment is smaller than the counterpart in the simulated datasets. The rationale lies in the fact that the practical non-idealities (e.g., hardware response and impairments) are not modeled in the simulated training datasets, which induces a sim-to-real gap for the cluster behaviour. To this end, a simulated channel dataset generated from a more practical channel model is required, which can characterize the cluster behaviour under non-idealities. Then, sim-to-real generalization gain of MCD in EG-CsiNet can be further enhanced.
\section{Conclusion}
In this paper, the environment-generalizability of deep learning-based CSI feedback is enhanced with intuitive physics interpretation. Firstly, the cross-environment distribution shift of the cluster-based channel is modeled, which comprises the distribution shift of the multi-cluster structure and the single-cluster response. Secondly, the physics-based distribution alignment is proposed to address the cross-environment distribution shift of the cluster-based channel, including multi-cluster decoupling and fine-grained alignment. Intuitively, the multi-cluster decoupling and fine-grained alignment can effectively address the distribution shift of the multi-cluster structure and the single-cluster response, respectively. Thirdly, the efficiency and robustness of the physics-based distribution alignment are enhanced. On the one hand, an efficient SVD-based multi-cluster decoupling algorithm is proposed to support real-time CSI feedback, which avoids the intermediate path-level parameter estimation. On the other hand, a hybrid criterion for noise-robust cluster number estimation is designed, which enables robust CSI feedback in various levels of downlink channel estimation error. Fourthly, the environment-generalizable EG-CsiNet is proposed as a universal learning framework for CSI feedback. Facilitated by the physics-based distribution alignment, the model training and inference of EG-CsiNet are designed to enhance generalization. Thanks
to the cluster-wise feedback manner, the proposed EG-CsiNet
can adaptively adjust the feedback overhead, and the model
parameters can be reduced as well. Comprehensive simulations and sim-to-real experiments are provided to justify the robust generalizability of the proposed EG-CsiNet over the SOTA, which can significantly reduce practical deployment costs of the deep learning-based CSI feedback.

\appendices
\section{Proof of \textbf{Theorem}~\ref{theo: optimal cluster}}
\label{appdix: theo 1 proof}
Based on the constraint $\mathcal{C}_{1}$ in \eqref{subeq: C1}, the rank-one cluster can be reformulated as $\mathbf{C}_{l}=\gamma_{l}\mathbf{x}_{l}\mathbf{y}_{l}^{H}$, where $\gamma_{l}>0$ and $\Vert\mathbf{x}_{l}\Vert_{2}=\Vert\mathbf{y}_{l}\Vert_{2}=1$. When $l\neq l^{\prime}$, orthogonality $\mathbf{x}_{l}^{H}\mathbf{x}_{l^{\prime}}=\mathbf{y}_{l}^{H}\mathbf{y}_{l^{\prime}}=0$ can be derived based on constraint $\mathcal{C}_{2}$ in \eqref{subeq: C2}. Thus, we can denote the summation $\overline{\mathbf{H}}=\sum_{l=1}^{\widehat{R}}\mathbf{C}_{l}=\sum_{l=1}^{\widehat{R}}\gamma_{l}\mathbf{x}_{l}\mathbf{y}_{l}^{H}$, which naturally yields a SVD formulation. Then, the optimization problem in \eqref{equ: cluster optimization} can be equivalently reformulated as a standard low-rank approximation problem
\begin{equation}
\label{equ: cluster optimization reformulate}
    \min_{\overline{\mathbf{H}}} \Vert\mathbf{H}-\overline{\mathbf{H}}\Vert_{F},\quad
    \text{s.t.}\quad \text{rank}(\overline{\mathbf{H}})=\widehat{R}.
\end{equation}
Based on the EYM theorem \cite{eckart1936approximation,mirsky1960symmetric}, the optimal $\overline{\mathbf{H}}^{\star}$ of \eqref{equ: cluster optimization reformulate} is yielded by $\overline{\mathbf{H}}^{\star}=\sum_{l=1}^{\widehat{R}}\sigma_{l}\mathbf{u}_{l}\mathbf{v}_{l}^{H}$, where $\sigma_{l}$ denotes the $l$th largest singular value of $\mathbf{H}$, and $\mathbf{u}_{l}$ and $\mathbf{v}_{l}$ denote the singular vectors. By comparing $\overline{\mathbf{H}}^{\star}$ and $\overline{\mathbf{H}}=\sum_{l=1}^{\widehat{R}}\mathbf{C}_{l}$, the optimal solution of \eqref{equ: cluster optimization} is yielded by $\mathbf{C}_{l}^{\star}=\sigma_{l}\mathbf{u}_{l}\mathbf{v}_{l}^{H}$, which completes the proof.  

\section{Proof of Angular-Delay Domain Peak-position Alignment via Phase Adjustment}

For the element-wise product $\mathbf{C}^{\prime}=\mathbf{S}\odot\mathbf{C}=\big(\text{conj}(\mathbf{w}^{(\rm a)}_{n^\star_1,n_2^\star})\otimes(\mathbf{w}_{m^\star}^{(\rm d)})^{T}\big)\odot\mathbf{\mathbf{C}}$, the angular-domain peak positions of $\mathbf{C}^{\prime}$ can be derived as 
\begin{equation}
    \label{equ: angular peak alignment}
    \begin{aligned}
    &(n_1^{(\rm aln)},n_2^{(\rm aln)})=\mathop{\arg\max}_{n_1,n_2}\left\{\Vert(\mathbf{w}_{n_1,n_2}^{(\rm a)})^{H}(\mathbf{S}\odot\mathbf{C})\Vert_{2}^{2}\right\}\\
    &\!=\!\mathop{\arg\max}_{n_1,n_2}\left\{\Vert(\mathbf{w}_{n_1,n_2}^{(\rm a)}\!\odot\!\mathbf{w}_{n^\star_1,n_2^\star}^{(\rm a)})^{H}(\mathbf{C}\odot(\mathbf{1}_{N_{\rm T}}\!\otimes\!(\mathbf{w}_{m^\star}^{(\rm d)})^{T}))\Vert_{2}^{2}\right\}\\
    &\!\overset{(a)}{=}\!\mathop{\arg\max}_{n_1,n_2}\left\{\Vert(\mathbf{w}_{n_1+n_1^\star,n_2+n_2^\star}^{(\rm a)})^{H}(\mathbf{C}\odot(\mathbf{1}_{N_{\rm T}}\otimes(\mathbf{w}_{m^\star}^{(\rm d)})^{T}))\Vert_{2}^{2}\right\}\\
    &\!\overset{(b)}{=}\!\mathop{\arg\max}_{n_1,n_2}\left\{\Vert(\mathbf{w}_{n_1+n_1^\star,n_2+n_2^\star}^{(\rm a)})^{H}\mathbf{C}\Vert_{2}^{2}\right\}\\
    &\!\overset{(c)}{=}\!(0,0),
    \end{aligned}
\end{equation}
where $\mathbf{1}_{n}$ denotes the $n$-dimensional all-ones vector. Here, subequation $\overset{(a)}{=}$ is held due to the property of the DFT codeword, i.e., 
\begin{equation}
    \begin{aligned}
    \mathbf{w}^{({\rm a},x)}_{n}\odot\mathbf{w}^{({\rm a},x)}_{m}&=\left[1,e^{{\rm j}2\pi \frac{n+m}{O_{x}N_{x}}},\ldots,e^{{\rm j}2\pi \frac{(n+m)(N_{x}-1)}{O_{x}N_{x}}}\right]^{T}\\
    &=\mathbf{w}^{({\rm a},x)}_{m+n}
    \end{aligned}
\end{equation}
for $x\in\{\rm{h,v}\}$, and the property of Kronecker products, i.e., 
\begin{equation}
    \label{equ: kronecker property}
    \begin{aligned}
    \mathbf{w}^{(\rm a)}_{n_1, m_1}\odot\mathbf{w}_{n_2,m_2}^{(\rm a)}&=(\mathbf{w}_{n_1}^{({\rm a, h})}\odot\mathbf{w}^{(\rm a,h)}_{n_2})\otimes(\mathbf{w}_{m_1}^{({\rm a, v})}\odot\mathbf{w}^{(\rm a,v)}_{m_2})\\
    &=\mathbf{w}^{({\rm a, h})}_{n_1+n_2}\otimes\mathbf{w}^{(\rm a, v)}_{m_1+m_2}\\
    &=\mathbf{w}^{(\rm a)}_{n_1+n_2, m_1+m_2}.
    \end{aligned}
\end{equation}
Subequation $\overset{(b)}{=}$ is held since the elements in $\mathbf{w}_{m^\star}^{(\rm d)}$ has unit modulus. Subequation $\overset{(c)}{=}$ is held based on \eqref{equ: peak position angular}. Based on \eqref{equ: angular peak alignment}, the angular-domain peak position is aligned to a fixed position $(0,0)$. Through similar derivations, it can be proved that the delay-domain peak position of $\mathbf{C}^\prime$ is 0. Therefore, the angular-delay domain peak positions of $\mathbf{C}$ can be aligned to a fixed position with the phase adjustment matrix $\mathbf{S}$. 
\label{appdix: phase adjustment proof}
\bibliographystyle{IEEEtran}
\bibliography{Ref}{}
\end{document}